\documentclass[twocolumn,english]{revtex4}
\usepackage{times}
\usepackage[T1]{fontenc}
\usepackage[latin1]{inputenc}
\usepackage{geometry}
\usepackage{amsmath}
\usepackage{graphicx}
\usepackage{amssymb}

\makeatletter

\providecommand{\tabularnewline}{\\}

\usepackage{babel}
\makeatother
\begin{document}

\title{Optical Feshbach resonances of Alkaline-Earth atoms in a 1D or 2D
optical lattice}

\author{Pascal Naidon$\,^{1}$}

\email{pascal.naidon@nist.gov}

\author{and Paul S. Julienne$\,^{1,2}$}

\affiliation{$\,^{1}$Atomic Physics Division and $\,^{2}$Joint Quantum Institute\\
National Institute of Standards and Technology, 100 Bureau Drive Stop
8423, Gaithersburg, Maryland 20899-8423, USA}

\date{\today}

\begin{abstract}
Motivated by a recent experiment by Zelevinsky \emph{et al.} {[}Phys.
Rev. Lett. \textbf{96}, 203201{]}, we present the theory for photoassociation
and optical Feshbach resonances of atoms confined in a tight one-dimensional
(1D) or two-dimensional (2D) optical lattice. In the case of an alkaline-earth
intercombination resonance, the narrow natural width of the line makes
it possible to observe clear manifestations of the dimensionality,
as well as some sensitivity to the scattering length of the atoms.
Among possible applications, a 2D lattice may be used to increase
the spectroscopic resolution by about one order of magnitude. Furthermore,
a 1D lattice induces a shift which provides a new way of determining
the strength of a resonance by spectroscopic measurements. 
\end{abstract}
\maketitle

\section{Introduction}

There is a growing experimental effort to study the properties of
ultracold alkaline-earth vapours. One of the reasons is that the narrow
intercombination resonance of alkaline-earth species (weakly coupling
$\,^{1}S_{0}$ and $\,^{3}P_{1}$ states) can be used to build optical
atomic clocks which could be more accurate than the current atomic
standard of time \cite{katori2003,ido2005,ludlow2006}. The Bose-condensation
of Ytterbium \cite{takasu2003}, whose atomic structure is close to
alkaline-earth species, has also raised the hope of condensing these
species. To reach these goals, a good knowledge, and possibly control,
of alkaline-earth atomic interactions are needed. Photoassociation,
the process of associating pairs of colliding atoms into excited bound
states by making them absorb a resonant photon, appears as the best
tool to characterize and control these interactions. Indeed, photoassociation
can be used as a spectroscopic tool to measure energy levels in excited
molecular states \cite{nagel2005,takasu2004,tojo2006} and characterize
ground-state atomic interactions \cite{michelson2005}. It can also
be regarded as an optical Feshbach resonance \cite{fedichev1996},
analogous to magnetic Feshbach resonances in alkali systems, making
it possible to alter these interactions \cite{ciurylo2005}. This
is particularly useful for alkaline-earth systems with no hyperfine
structure, where magnetic Feshbach resonance is not possible.

Photoassociation near the intercombination resonance is characterized
by narrow lines which are sensitive to effects like recoil shifts
(due to the momentum of the absorbed photon) and Doppler broadening.
These effects can be eliminated by strongly confining the atoms in
the direction of propagation of the photoassociation light. This has
been demonstrated by trapping the atoms in tight optical lattices
\cite{ido2003,takamoto2003,ludlow2006,zelevinsky2006}. The wave length
of the lattice can be adjusted to its {}``magic'' value \cite{ido2003},
so that atoms in the ground state and the excited state feel the same
trapping potential.

In turn, it is known that strong confinement may affect the collisional
properties of the atoms \cite{olshanii1998,petrov2001}. We can therefore
expect photoassociation to be different in an optical lattice than
it is in free space. The purpose of this paper is to investigate how
photoassociation spectroscopy and optical Feshbach resonances are
affected by a one-dimensional (1D) or two-dimensional (2D) optical
lattice. In Section~\ref{sec:Effects-of-the-Lattice}, we present
the general theory of resonant collisions in a 1D and 2D optical lattice,
building upon the works of D. S. Petrov and G. V. Shlyapnikov \cite{petrov2001}
and M. Olshanii \cite{olshanii1998}. In section~\ref{sec:AlkalineEarth},
we apply the theory to the case of alkaline-earth species. In section~\ref{sec:Conclusion},
we conclude on the possible effects and uses of alkaline-earth photoassociation
in optical lattices.

\section{Optical resonances in a 1d or 2D lattice\label{sec:Effects-of-the-Lattice}}

An optical lattice is obtained using a standing wave laser which creates
a sinusoidal periodic trap. Each cell of this lattice confines the
atoms in either a pancake-shaped cloud (for 1D lattices) or cigar-shaped
cloud (for 2D lattices). For this reason, we will refer to 1D lattices
as a 2D (pancake) geometry, and to 2D lattices as 1D (cigar) geometry.
We will assume that there is little transfer between each cloud, so
that we can treat each one independently. The atoms in each cloud
will be assumed free to move in the directions which are not confined
by the lattice, ie, the transverse directions $x,y$ for a 1D lattice,
and the axial direction $z$ for a 2D lattice. We will also describe
the confinement induced by the lattice near the centre of each cell
by a harmonic potential, which is valid if the lattice is tight or
strong enough.

To describe optical resonances under such confinement, we use the
theories of collisions in 1D or 2D geometries of Refs.~\cite{olshanii1998,petrov2001}.
We express these theories in paragraph \ref{sub:Collisions-in-confinement}
in terms of an energy-dependent scattering length $a(k)$, which is
essentially the $K$-matrix for 3D collisions. The original theories
were formulated in terms of the usual (energy-independent) scattering
length, but taking into account the energy-dependence extends the
range of validity of the theories \cite{naidon2006}, and is necessary
in the case of resonant collisions. We then give the expression for
$a(k)$ in the case of resonant collision in paragraph \ref{sub:Optical-Feshbach-resonance}.
Using this expression, we deduce the collisional and photoassociation
rates in 1D and 2D geometries.

Similar methods have been used to treat specific cases: a $K$-matrix
approach was used in Ref. \cite{wouter2003} to describe bosonic resonant
collisions in a 2D geometry, a multi-channel $K$-matrix approach
was used in Ref. \cite{granger2004} to describe collisions of spin-polarized
fermions in a 1D geometry, and a renormalization approach was used
in Refs. \cite{yurovsky2005,yurovsky2006} to describe resonant collisions
of bosons in a 1D geometry. All these works relate the effective 1D
or 2D scattering properties to the free-space properties. Also closely
related are the constructions of pseudo-potentials which can treat
confined systems \cite{bolda2003,kanjilal2004,stock2005} and pseudo-potentials
in low-dimensional systems \cite{kanjilal2004,kanjilal2006}.

\subsection{Collisions in a 1D or 2D confined geometry\label{sub:Collisions-in-confinement}}

Let us briefly review here the method of Refs.~\cite{olshanii1998,petrov2001}.
The stationary states $\psi(\mathbf{r})$ for the relative motion
of two atoms confined by a harmonic potential $V(r)$ are given by
the Schr\"odinger equation\begin{equation}
\left[-\frac{\hbar^{2}}{2\mu}\nabla_{\vec{r}}^{2}+U(r)+V(\vec{r})\right]\psi(\vec{r})=E\psi(\vec{r})\label{eq:Schrodinger}\end{equation}
where $\vec{r}=(x,y,z)$ is the relative coordinate between the two
atoms with length $r=\vert\vec{r}\vert$, $\mu$ is their reduced
mass, $U(r)$ is their interaction potential, $E$ is their relative
motion energy. In a 2D geometry (1D confinement) the trapping potential
is:\[
V(\vec{r})\equiv V_{2D}(\vec{r})=\frac{1}{2}\mu\Omega^{2}z^{2}\]
where $\Omega$ is the frequency of the trap. In a 1D geometry (2D
confinement), it is:\[
V(\vec{r})\equiv V_{1D}(\vec{r})=\frac{1}{2}\mu\Omega^{2}\vec{\rho}^{\,2}\]
 with $\vec{\rho}=(x,y)$.

At separation $r$ larger than the range $r_{0}$ of the interaction
potential $U$, the stationary states can be written as a linear combination
of products of a harmonic oscillator eigenstate in the confined direction
and a free wave in the weak direction. We choose the conventional
basis where each stationary state in the weak direction is asymptotically
the sum of an incident plane wave and a scattered wave (which is assumed
isotropic for low energies). In 2D and 1D, we have respectively:\begin{equation}
\psi_{\nu,\vec{q}}^{2D}(\vec{r})\xrightarrow[r>r_{0}]{}\varphi_{\nu}(z)e^{i\vec{q}\cdot\vec{\rho}}+\sum_{\nu}f_{\nu,\nu^{\prime}}^{E}\varphi_{\nu^{\prime}}(z)\sqrt{\frac{i}{8\pi q^{\prime}\rho}}e^{iq^{\prime}\rho}\label{eq:IncidentAndScatteredWave2D}\end{equation}
\begin{equation}
\psi_{n,m,p}^{1D}(\vec{r})\mathop{\to}_{r>r_{0}}\phi_{nm}(\rho)e^{ipz}+\sum_{n^{\prime}m^{\prime}}f_{nm,n^{\prime}m^{\prime}}^{E}\phi_{n^{\prime}m^{\prime}}(\rho)e^{ip^{\prime}\vert z\vert}\label{eq:IncidentAndScatteredWave1D}\end{equation}
where $\varphi_{\nu}$, $\phi_{nm}$ are respectively the unit-normalized
1D (2D) harmonic oscillator wave function of vibration index $\nu$
(principal number $n$ and angular momentum number $m$), $p$ is
the wave number of the incoming 1D wave, and $\vec{q}$ is the wave
vector of the incoming 2D plane wave with wave number $q$. For each
state, the total energy $E$ is composed of the oscillator energy
in the confined direction, and the free particle energy in the weak
direction:\begin{eqnarray}
E_{2D} & = & \begin{array}{c}
\hbar\Omega(\frac{1}{2}+\nu)\end{array}+\frac{\hbar^{2}\vec{q}^{\,2}}{2\mu}\label{eq:Energy2D}\\
 & = & \begin{array}{c}
\hbar\Omega(\frac{1}{2}+\nu^{\prime})\end{array}+\frac{\hbar^{2}q^{\prime2}}{2\mu}\nonumber \end{eqnarray}
\begin{eqnarray}
E_{1D} & = & \begin{array}{c}
\hbar\Omega(1+2n+\vert m\vert)\end{array}+\frac{\hbar^{2}p^{2}}{2\mu}\label{eq:Energy1D}\\
 & = & \begin{array}{c}
\hbar\Omega(1+2n^{\prime}+\vert m^{\prime}\vert)\end{array}+\frac{\hbar^{2}p^{\prime2}}{2\mu}\nonumber \end{eqnarray}
which defines the wave numbers $p$ and $q$ for the incoming plane
wave, as well as $p^{\prime}$ and $q^{\prime}$ for the scattered
wave. The basis of eigenstates~(\ref{eq:IncidentAndScatteredWave1D})
and (\ref{eq:IncidentAndScatteredWave2D}) define respectively the
effective 2D (1D) scattering amplitude $f_{\nu,\nu^{\prime}}^{E}$
($f_{nm,n^{\prime}m^{\prime}}^{E}$) for two atoms colliding with
relative energy $E$ in the initial state $\varphi_{\nu}$ ($\phi_{nm}$)
to end up in the final state $\varphi_{\nu^{\prime}}$ ($\phi_{n^{\prime}m^{\prime}}$).

At short separations, one can neglect the confining trap in Eq.~(\ref{eq:Schrodinger})
and the collision recovers the aspects of a usual 3D (free space)
collision at energy $E=\frac{\hbar^{2}k^{2}}{2\mu}$. As a result,
as long as $kr_{0}\lesssim1$, there is a region \cite{naidon2006}
where the the wave function is proportional to the usual 3D \emph{s}-wave
scattering wave function\begin{equation}
\psi(\mathbf{r})\propto\frac{\sin kr}{kr}-a(k)\frac{\cos kr}{r},\label{eq:3DWaveFunction}\end{equation}
where $a(k)$ is the energy-dependent $s$-wave scattering length
introduced in Ref.~\cite{blume2002,bolda2002}. At very low energy,
the scattering length $a(0)$ is simply the usual scattering length
$a$. More generally, it is related to the 3D elastic scattering matrix
element $S(k)$ by:\begin{equation}
a(k)=\frac{1}{ik}\frac{1-S(k)}{1+S(k)}\label{eq:DefinitionEffectiveScatteringLength}\end{equation}

By matching (\ref{eq:IncidentAndScatteredWave1D}) or (\ref{eq:IncidentAndScatteredWave2D})
with (\ref{eq:3DWaveFunction}), one finds the relation between the
effective 1D or 2D scattering amplitude and the scattering length
\cite{bergeman2003a,petrov2001}:\begin{multline}
f_{\nu,\nu^{\prime}}^{E}=-4\pi a(k)\varphi_{\nu^{\prime}}(0)\varphi_{\nu}(0)\,\eta_{2D}\\
\times\theta\left(E-\hbar\Omega(\begin{array}{c}
\!\frac{1}{2}\end{array}\!\!+\nu^{\prime})\right)\theta\left(E-\hbar\Omega(\!\begin{array}{c}
\frac{1}{2}\!\!\end{array}+\nu)\right)\label{eq:2DAmplitude}\end{multline}
\begin{multline}
f_{nm,n^{\prime}m^{\prime}}^{E}=\frac{4\pi a(k)}{2ip^{\prime}}\phi_{n^{\prime}m^{\prime}}(0)\phi_{nm}(0)\,\eta_{1D}\\
\times\theta\left(E-\hbar\Omega(1+2n^{\prime})\right)\theta\left(E-\hbar\Omega(1+2n)\right)\label{eq:2DAmplitude}\end{multline}
where $\theta$ is the Heavyside function and\begin{equation}
\eta_{2D}=\frac{1}{1+\frac{a(k)}{a_{0}}\frac{1}{\sqrt{\pi}}w(\frac{\nu}{2}+(\frac{qa_{0}}{2})^{2})}\label{eq:eta2D}\end{equation}
\begin{equation}
\eta_{1D}=\frac{1}{1+\frac{a(k)}{a_{0}}\zeta(\frac{1}{2},-n-(\frac{pa_{0}}{2})^{2})}\label{eq:eta1D}\end{equation}
where $a_{0}=\sqrt{\hbar/\mu\Omega}$ is the harmonic oscillator length,
$w(x)$ is a function introduced in Ref.~\cite{petrov2001}, and
$\zeta(\alpha,x)$ is the Hurwitz zeta function (as defined in Ref.~\cite{bergeman2003a}).
These functions have the following properties:\begin{equation}
\begin{array}{ccll}
\frac{1}{\sqrt{\pi}}w(x) & \sim & 2i\sqrt{x} & \textrm{for }x\gg1\\
 & \sim & i\sqrt{\pi} & \textrm{for }0.1<x\lesssim1\\
 & \sim & i\sqrt{\pi}+\frac{1}{\sqrt{\pi}}\ln(B/2\pi x)\quad & \textrm{for }x\ll1\end{array}\label{eq:wProperties}\end{equation}
\begin{equation}
\begin{array}{ccll}
\zeta(\frac{1}{2},-x) & \sim & 2i\sqrt{x} & \textrm{for }x\gg1\\
 & = & \zeta(\frac{1}{2},1-x)+i/\sqrt{x} & \textrm{for }0<x<1\\
 & \sim & \zeta(\frac{1}{2})+i/\sqrt{x}\quad & \textrm{for }x\ll1\end{array}\label{eq:zetaProperties}\end{equation}
where $B\approx0.915$ and $\zeta(\frac{1}{2})\approx-1.46$. Note
that, due to the properties of $\varphi_{\nu}$ and $\phi_{nm}$ \cite{moore2004},
$f_{\nu,\nu^{\prime}}^{E}$ is non-zero only when both $\nu$ and
$\nu^{\prime}$ are even, and $f_{nm,n^{\prime}m^{\prime}}^{E}$ is
non-zero only when $m=m^{\prime}=0$.

\subsection{Optical Feshbach resonances in free space\label{sub:Optical-Feshbach-resonance}}

Usual photoassociation in free space can be modelled by a multichannel
scattering theory \cite{bohn1999}. In the case of a photoassociation
resonance with an isolated molecular state, one finds simple expressions
for the scattering properties of atoms under the influence of the
laser light. For instance, the \emph{s}-wave elastic scattering matrix
element reads:\begin{equation}
S(k)=S_{\textrm{bg}}(k)\left(1-\frac{i\Gamma(k)}{E-E_{0}+\frac{i}{2}\left(\Gamma(k)+\gamma\right)}\right)\label{eq:PhotoassociationSMatrix}\end{equation}

Here, $S_{\textrm{bg}}(k)$ is the $s$-wave elastic scattering matrix
element in the absence of light, $E=\frac{\hbar^{2}k^{2}}{2\mu}$
is the collision energy, $E_{0}$ is the position of the resonance
(with respect to the ground-state threshold), $\Gamma(k)$ is the
stimulated width of the resonance, and $\gamma$ is the natural width
of the resonant molecular state. The stimulated width $\Gamma(k)$
is due to the coupling by the laser between the incoming scattering
state and the resonant molecular state, and is proportional to the
intensity of the laser light. The natural width $\gamma$ is due to
the decay of the resonant molecular state by spontaneous emission.
Because of this natural width, the scattering matrix element is no
longer unitary $(\vert S(k)\vert^{2}<1)$, meaning that the collision
is no longer elastic. This accounts for the fact that some photoassociated
pairs of atoms can deexcite by spontaneous emission, and do not return
to their initial state. In most cases, these pairs of atoms gain enough
energy to leave the trap where the system is confined, or form bound
molecules \cite{fioretti1998}, so that photoassociation is observed
as a loss from the initial atomic cloud. The observed loss serves
as a basis for photoassociative spectroscopy.

A magnetic Feshbach resonance is described by a formally similar theory,
where spontaneous emission is generally negligible ($\gamma=0$).The
following theory is therefore applicable to a magnetic Feshbach resonance
by setting $\gamma\to0$ and $\gamma l_{\textrm{opt}}\to\kappa$,
where $\kappa$ is the strength of the magnetic resonance.

From the expression (\ref{eq:PhotoassociationSMatrix}) of the scattering
matrix element, one can derive the energy-dependent scattering length
(\ref{eq:DefinitionEffectiveScatteringLength}) near a photoassociation
resonance:\begin{equation}
a(k)=\frac{a_{\textrm{bg}}(k)+l_{\textrm{opt}}(k)\left(\frac{E-E_{0}}{\gamma}+\frac{i}{2}\right)^{-1}}{1-k^{2}a_{\textrm{bg}}(k)l_{\textrm{opt}}(k)\left(\frac{E-E_{0}}{\gamma}+\frac{i}{2}\right)^{-1}}\label{eq:Photoassociationak}\end{equation}

where $a_{\textrm{bg}}(k)=\frac{1}{ik}(1-S_{\textrm{bg}}(k))/(1+S_{\textrm{bg}}(k))$
is the background energy-dependent scattering length in the absence
of light, and the optical length $l_{\textrm{opt}}(k)$ is defined
by \cite{ciurylo2005}:\begin{equation}
l_{\textrm{opt}}(k)=\frac{\Gamma(k)}{2\gamma k}.\label{eq:OpticalLength}\end{equation}

\paragraph*{Wigner's threshold law\protect \\
}

For sufficiently small collisional energies, the incoming scattering
state obeys Wigner's threshold law. Namely, the energy-dependent background
scattering length $a_{\textrm{bg}}(k)$ becomes constant\begin{equation}
a_{\textrm{bg}}(k)\approx a_{\textrm{bg}},\label{eq:Wigner1}\end{equation}
 when $\frac{1}{2}k^{2}a_{\textrm{bg}}r_{e}\ll1$, where $r_{e}$
is the effective range \cite{bethe1949} of the background interaction
(usually on the order of $r_{0}$). Here, $a_{\textrm{bg}}$ denotes
the usual zero-energy background scattering length $a_{\textrm{bg}}(0)$.
Similarly, for $kr_{e}\ll kr_{C}\ll1$, where $r_{C}$ is the Condon
point of the photoassociation transition, the stimulated width is
proportional to the collisional momentum, and the optical length becomes
a constant:\begin{equation}
l_{\textrm{opt}}(k)\approx l_{\textrm{opt}}.\label{eq:Wigner2}\end{equation}
This constant $l_{\textrm{opt}}$ characterizes the strength of the
resonance at low energy. 

In many cases, $ka_{\textrm{bg}}$ and $kl_{\textrm{opt}}$ are small,
so that Eq.~(\ref{eq:Photoassociationak}) simplifies to:\begin{equation}
a(k)\approx a_{\textrm{bg}}+l_{\textrm{opt}}\left(\frac{E-E_{0}}{\gamma}+\frac{i}{2}\right)^{-1}\label{eq:ModifiedScatteringLength}\end{equation}
which is the usual expression for the scattering length of the atoms
modified by the radiative coupling.

\subsection{Resonance in a 3D geometry}

In a 3D (almost free space) geometry, the number of (collision or
photoassociation) events per unit of time is given by:\begin{equation}
\int\mathcal{K}^{3D}\left(n^{3D}(x,y,z)\right)^{2}dxdydz\label{eq:3DrateEquation}\end{equation}
where $n^{3D}$ is the density of atoms and $\mathcal{K}^{3D}$ is
the usual 3D (collision or photoassociation) rate, which can be calculated
from the scattering properties of the atoms, such as the scattering
matrix element (\ref{eq:PhotoassociationSMatrix}). They are given
respectively by:\begin{equation}
\begin{array}{clc}
\mathcal{K}_{\textrm{col}}^{3D} & = & {\displaystyle \frac{\pi\hbar}{\mu}}\left\langle \frac{\vert1-S(k)\vert^{2}}{k}\right\rangle _{\vec{k}}\\
\mathcal{K}_{\textrm{pa}}^{3D} & = & {\displaystyle \frac{\pi\hbar}{\mu}}\left\langle \frac{1-\vert S(k)\vert^{2}}{k}\right\rangle _{\vec{k}}\end{array}\label{eq:General3Drates}\end{equation}
where\[
\Big\langle\dots\Big\rangle_{\vec{k}}=\int d^{3}\vec{k}\left(\dots\right)\mathcal{P}(\vec{k})\]
 denotes statistical average over all the possible collision wave
vectors $\vec{k}$, and $\mathcal{P}(\vec{k})$ is the wave vector
distribution normalised to 1. One can see in (\ref{eq:General3Drates})
that the photoassociation rate follows from the non-unitarity of the
scattering matrix, $\vert S(k)\vert^{2}<1$. From Eq.~(\ref{eq:PhotoassociationSMatrix}),
we also note that the photoassociation rate does not depend directly
on the background scattering length $a_{\textrm{bg}}$. There is,
of course, an indirect dependence through $\Gamma$ or $l_{\textrm{opt}}$
which contain a Franck-Condon factor involving the ground-state wave
function, whose nodal structure depends on the background scattering
length. This indirect dependence has been used to infer the background
scattering length from photoassociation spectra \cite{michelson2005,yasuda2006}.

\subsection{Resonance in a 2D geometry}

In a 2D geometry, the number of (collision or photoassociation) events
per unit of time is given by:\begin{equation}
\int\mathcal{K}^{2D}\left(n^{2D}(x,y)\right)^{2}dxdy\label{eq:2DrateEquation}\end{equation}
where $n^{2D}$ is the 2D density of atoms, and the rates of collision
or photoassociation $\mathcal{K}^{2D}$ are related to the quasi-2D
scattering matrix element $S_{\nu\nu^{\prime}}^{E}=\delta_{\nu\nu^{\prime}}+\frac{i}{2}f_{\nu\nu^{\prime}}^{E}$:

\begin{equation}
\begin{array}{cll}
\mathcal{K}_{\textrm{col}}^{2D} & = & {\displaystyle \frac{\hbar}{\mu}}\left\langle {\displaystyle \sum_{\nu^{\prime}}}\vert\delta_{\nu\nu^{\prime}}-S_{\nu\nu^{\prime}}^{E}\vert^{2}\right\rangle _{\nu,\vec{q}}\\
\mathcal{K}_{\textrm{pa}}^{2D} & = & {\displaystyle \frac{\hbar}{\mu}}\left\langle {\displaystyle \sum_{\nu^{\prime}}}\left(\delta_{\nu\nu^{\prime}}-\vert S_{\nu\nu^{\prime}}^{E}\vert^{2}\right)\right\rangle _{\nu,\vec{q}}\end{array}\label{eq:General2Drates}\end{equation}

In these expressions, a sum is performed over all final states $\nu^{\prime}$,
and the average

\[
\Big\langle\dots\Big\rangle_{\nu,\vec{q}}=\int\!\! d^{2}\vec{q}\sum_{\nu\textrm{ even}}\left(\dots\right)\mathcal{P}(\nu,\vec{q})\]
denotes the statistical average over all initial states of relative
motion. The statistical distribution $\mathcal{P}(\nu,\vec{q})$ is
normalised to unity: $\int\! d^{2}\vec{q}\;\sum_{\nu}\mathcal{P}(\nu,\vec{q})=1$. 

From Eq. (\ref{eq:2DAmplitude}) and the properties of $w$, one can
derive:\begin{eqnarray}
\mathcal{K}_{\textrm{col}}^{2D} & = & \frac{4\hbar}{\mu}\sqrt{\pi}\left\langle \frac{\frac{(\nu-1)!!}{\nu!!}\textrm{Im}\left[\frac{1}{\sqrt{\pi}}w\left(\frac{\nu}{2}+\frac{(qa_{0})^{2}}{4}\right)\right]}{\left|\frac{a_{0}}{a(k)}+\frac{1}{\sqrt{\pi}}w\left(\frac{\nu}{2}+\frac{(qa_{0})^{2}}{4}\right)\right|^{2}}\right\rangle _{\nu,\vec{q}}\label{eq:2DCollisionRate}\\
\mathcal{K}_{\textrm{pa}}^{2D} & = & \frac{4\hbar}{\mu}\sqrt{\pi}\left\langle \frac{\frac{(\nu-1)!!}{\nu!!}\textrm{Im}\left[\frac{a_{0}}{a(k)}\right]}{\left|\frac{a_{0}}{a(k)}+\frac{1}{\sqrt{\pi}}w\left(\frac{\nu}{2}+\frac{(qa_{0})^{2}}{4}\right)\right|^{2}}\right\rangle _{\nu,\vec{q}}\label{eq:2DPhotoassociationRate}\end{eqnarray}
where we recall that $k^{2}=\frac{2\nu+1}{a_{0}^{2}}+q^{2}$ and $n!!$
is the double factorial of $n$.

\subsection{Resonance in a 1D geometry}

In a 1D geometry, the number of (collision or photoassociation) events
per unit of time is given by:\begin{equation}
\int\mathcal{K}^{1D}\left(n^{1D}(z)\right)^{2}dz\label{eq:1DrateEquation}\end{equation}
where $n^{1D}$ is the 1D density of atoms, and the rates of collision
or photoassociation $\mathcal{K}^{1D}$ are related to the quasi-1D
scattering matrix element $S_{nm,n^{\prime}m^{\prime}}^{E}=\delta_{nn^{\prime}}\delta_{mm^{\prime}}+2f_{nm,n^{\prime}m^{\prime}}^{E}$:

\begin{equation}
\begin{array}{cll}
\mathcal{K}_{\textrm{col}}^{1D} & = & {\displaystyle \frac{\hbar}{2\mu}}\left\langle {\displaystyle \sum_{n^{\prime}m^{\prime}}}\vert\delta_{nn^{\prime}}\delta_{mm^{\prime}}-S_{nm,n^{\prime}m^{\prime}}^{E}\vert^{2}p^{\prime}\right\rangle _{n,m,p}\\
\mathcal{K}_{\textrm{pa}}^{1D} & = & {\displaystyle \frac{\hbar}{2\mu}}\left\langle {\displaystyle \sum_{n^{\prime}m^{\prime}}}\left(\delta_{nn^{\prime}}\delta_{mm^{\prime}}-\vert S_{nm,n^{\prime}m^{\prime}}^{E}\vert^{2}\right)p^{\prime}\right\rangle _{n,m,p}\end{array}\label{eq:General1Drates}\end{equation}

In these expressions, a sum is performed over all final states $n^{\prime},m^{\prime}$,
and the average

\[
\Big\langle\dots\Big\rangle_{n,m,p}=\int\!\! dp\,\sum_{n,m}\left(\dots\right)\mathcal{P}(n,m,p)\]
denotes the statistical average over all initial states of relative
motion. The statistical distribution $\mathcal{P}(n,m,p)$ is normalised
to unity: $\int\! dp\,\sum_{n,m}\mathcal{P}(n,m,p)=1$. 

From Eq. (\ref{eq:2DAmplitude}) and the properties of $\zeta$, one
can derive:\begin{eqnarray}
\mathcal{K}_{\textrm{col}}^{1D} & = & \frac{4\hbar}{\mu a_{0}}\left\langle \frac{\textrm{Im}\left[\zeta\left(\frac{1}{2},-n-\frac{(pa_{0})^{2}}{4}\right)\right]}{\left|\frac{a_{0}}{a(k)}+\zeta\left(\frac{1}{2},-n-\frac{(pa_{0})^{2}}{4}\right)\right|^{2}}\delta_{m,0}\right\rangle _{n,m,p}\label{eq:1DCollisionRate}\\
\mathcal{K}_{\textrm{pa}}^{1D} & = & \frac{4\hbar}{\mu a_{0}}\left\langle \frac{\textrm{Im}\frac{a_{0}}{a(k)}}{\left|\frac{a_{0}}{a(k)}+\zeta\left(\frac{1}{2},-n-\frac{(pa_{0})^{2}}{4}\right)\right|^{2}}\delta_{m,0}\right\rangle _{n,m,p}\label{eq:1DPhotoassociationRate}\end{eqnarray}
where we recall that $k^{2}=2\frac{1+2n+\vert m\vert}{a_{0}^{2}}+p^{2}$.

\subsection{Comparison of 2D/1D and 3D theories at thermal equilibrium\label{sub:Comparison-of-2D-1D}}

The comparison of 2D or 1D rates with the more familiar 3D rates is
not direct. It is more easily achieved at thermal equilibrium. First,
we can evaluate these rates assuming a Boltzmann distribution in all
cases:\begin{eqnarray*}
\mathcal{P}(\vec{k}) & = & \alpha^{3/2}\exp\left(-\frac{1}{k_{B}T}\frac{\hbar^{2}k^{2}}{2\mu}\right)\\
\mathcal{P}(\nu,\vec{q}) & = & \alpha\beta\exp\left(-\frac{\hbar\Omega\nu+\frac{\hbar^{2}q^{2}}{2\mu}}{k_{B}T}\right)\\
\mathcal{P}(n,m,p) & = & \alpha^{1/2}\beta^{2}\exp\left(-\frac{\hbar\Omega(2n+\vert m\vert)+\frac{\hbar^{2}p^{2}}{2\mu}}{k_{B}T}\right)\end{eqnarray*}
where $\alpha=\frac{\hbar^{2}}{2\pi\mu k_{B}T}$ and $\beta=1-e^{-\frac{\hbar\Omega}{k_{B}T}}$.
Secondly, in a strongly confined gas at temperature $T$, the density
profile in the confined direction can be assumed to be {}``frozen''
in the Boltzman profile (at least, for a certain time):\begin{eqnarray*}
n^{3D}(x,y,z,t) & \approx & n^{2D}(x,y,t)\times f_{T}(z)\\
n^{3D}(x,y,z,t) & \approx & n^{1D}(z,t)\times f_{T}(x)f_{T}(y)\end{eqnarray*}
with the Boltzmann profile:\[
f_{T}(x)=\frac{1}{\sqrt{\pi}L}e^{-\left(\frac{x}{L}\right)^{2}}\]
where $L=a_{0}/\sqrt{2\tanh(\hbar\Omega/2k_{B}T)}$. Note that a thermal
distribution of independent distinguishable atoms is assumed here.
Then we can state that the number of events per unit of time (\ref{eq:2DrateEquation})
or (\ref{eq:1DrateEquation}) is given by the 3D formula (\ref{eq:3DrateEquation})
with an effective 3D rate $\bar{\mathcal{K}}^{2D}$, or $\bar{\mathcal{K}}^{1D}$
respectively. Integrating out the profile in the confined direction,
we can find the relation between the 2D/1D rate and the effective
3D rate:\begin{eqnarray*}
\bar{\mathcal{K}}^{2D} & = & \sqrt{2\pi}L\;\times\mathcal{K}^{2D},\\
\bar{\mathcal{K}}^{1D} & = & \left(\sqrt{2\pi}L\right)^{2}\times\mathcal{K}^{1D}.\end{eqnarray*}
The effective 3D rate thus calculated from the 2D or 1D theory can
be compared with the usual 3D rate (\ref{eq:General3Drates}). In
the following, we consider three regimes identified in Ref.~ \cite{petrov2001},
corresponding to the different regions (\ref{eq:wProperties}) of
the function $w$ or $\zeta$.

\subsubsection*{Free space regime $k_{B}T\gg\hbar\Omega$}

In this limit, a large number of incoming $\nu$, or $n,m$ are involved:
the states of relative motion in the confined direction form a quasi-continuum
as in free space. Thus, we expect to retrieve the free-space results.
Indeed, both in the 2D and 1D theory, the effective 3D rates simplify
to:\begin{eqnarray}
\bar{\mathcal{K}}_{\textrm{col}}^{3D} & \!\!=\!\! & \left\langle \frac{\frac{2h}{\mu}\times k\times\vert a(k)\vert^{2}}{\left|1+ika(k)\right|^{2}}\right\rangle _{\vec{k}}\label{eq:FreeSpaceColRate}\\
\bar{\mathcal{K}}_{\textrm{pa}}^{3D} & \!\!=\!\! & \left\langle \frac{\frac{2h}{\mu}\times\textrm{Im}(-a(k))}{\left|1+ika(k)\right|^{2}}\right\rangle _{\vec{k}}\label{eq:FreeSpacePARate}\end{eqnarray}
 One can check that these expressions are exactly the same as the
usual 3D rates (\ref{eq:General3Drates}). In particular, if the Wigner's
threshold laws (\ref{eq:Wigner1}-\ref{eq:Wigner2}) apply (\emph{ie}
if the temperature is not too high), the photoassociation rate is
given by the standard expression:\begin{eqnarray}
\mathcal{K}_{\textrm{pa}}^{3D} & \!\!=\!\! & \int_{0}^{\infty}\!\!\!\frac{\frac{2h}{\sqrt{\pi}\mu}l_{\textrm{opt}}\times e^{-\frac{E}{k_{B}T}}}{\left(\frac{E-E_{0}}{\gamma}\right)^{2}+\left(\frac{1}{2}+kl_{\textrm{opt}}\right)^{2}}\frac{\sqrt{E}dE}{(k_{B}T)^{3/2}}\label{eq:FreeSpaceRateWignerLimit}\end{eqnarray}
where $E=\frac{\hbar^{2}k^{2}}{2\mu}$. The term $kl_{\textrm{opt}}$
is responsible for the so-called power-broadening by the laser.

\subsubsection*{Confinement-dominated regime $0.1\hbar\Omega\lesssim k_{B}T\lesssim\hbar\Omega$}

In this regime, very few incoming states in the confined direction
contribute to the collision. We can make the approximation that only
the $\nu=0$, or $n,m=0$ incoming states make a significant contribution.
In a 2D geometry, the effective 3D rates become: \begin{eqnarray}
\bar{\mathcal{K}}_{\textrm{col}}^{2D} & \!\!=\!\! & \lambda\left\langle \frac{\frac{2h}{\mu}\times\frac{\sqrt{\pi}}{a_{0}}\times\vert a(k)\vert^{2}}{\left|1+i\frac{\sqrt{\pi}}{a_{0}}a(k)\right|^{2}}\right\rangle _{\vec{q}}\label{eq:2DConfinedColRate}\\
\bar{\mathcal{K}}_{\textrm{pa}}^{2D} & \!\!=\!\! & \lambda\left\langle \frac{\frac{2h}{\mu}\times\textrm{Im}\left[-a(k)\right]}{\left|1+i\frac{\sqrt{\pi}}{a_{0}}a(k)\right|^{2}}\right\rangle _{\vec{q}}\label{eq:2DConfinedPARate}\end{eqnarray}
where $\Big\langle\dots\Big\rangle_{\vec{q}}=\int d^{2}\vec{q}\left(\dots\right)\mathcal{P}(\vec{q})$
is the statistical average over all 2D momenta, $\mathcal{P}(\vec{q})=\frac{\hbar^{2}}{2\pi\mu k_{B}T}\exp\left(-\frac{\frac{\hbar^{2}q^{2}}{2\mu}}{k_{B}T}\right)$
is the 2D Boltzmann distribution, $k=\sqrt{a_{0}^{-2}+q^{2}}$, and
$\lambda=\sqrt{1-e^{-2\frac{\hbar\Omega}{k_{B}T}}}$ is a correction
factor close to 1. Comparison with (\ref{eq:FreeSpaceColRate}-\ref{eq:FreeSpacePARate})
shows that the 1D confinement brings two main effects:

\begin{itemize}
\item the 3D momentum integral becomes a 2D integral
\item the momenta involved in the integral now have a lower bound given
by the {}``zero-point momentum'' $\hbar/a_{0}$. Note that $k$
is replaced by $\sqrt{a_{0}^{-2}+q^{2}}$ in $a(k)$, and by $\sqrt{\pi}/a_{0}$
elsewhere.
\end{itemize}
~ \\
\\
In a 1D geometry, the effective 3D rates become: \begin{eqnarray}
\!\!\!\!\!\!\bar{\mathcal{K}}_{\textrm{col}}^{1D} & \!\!\!=\!\! & \lambda^{2}\!\left\langle \!\frac{\frac{2h}{\mu}\times\frac{2}{pa_{0}^{2}}\times\vert a(k)\vert^{2}}{\left|1+\frac{a(k)}{a_{0}}\zeta(\frac{1}{2},1\!\!-\!\!(\frac{pa_{0}}{2})^{2})+i\frac{2}{pa_{0}^{2}}a(k)\right|^{2}}\!\right\rangle _{p}\label{eq:1DConfinedColRate}\\
\!\!\!\!\!\!\bar{\mathcal{K}}_{\textrm{pa}}^{1D} & \!\!\!=\!\! & \lambda^{2}\!\left\langle \!\frac{\frac{2h}{\mu}\times\textrm{Im}\left[-a(k)\right]}{\left|1+\frac{a(k)}{a_{0}}\zeta(\frac{1}{2},1\!\!-\!\!(\frac{pa_{0}}{2})^{2})+i\frac{2}{pa_{0}^{2}}a(k)\right|^{2}}\!\right\rangle _{p}\label{eq:1DConfinedPARate}\end{eqnarray}
where $\Big\langle\dots\Big\rangle_{p}=\int_{-\infty}^{\infty}dp\left(\dots\right)\mathcal{P}(p)$
is the statistical average over all 1D momenta, $\mathcal{P}(p)=\left(\frac{\hbar^{2}}{2\pi\mu k_{B}T}\right)^{1/2}\exp\left(-\frac{\frac{\hbar^{2}p^{2}}{2\mu}}{k_{B}T}\right)$
is the 1D Boltzmann distribution, and $k=\sqrt{2a_{0}^{-2}+p^{2}}$.
Comparison with (\ref{eq:FreeSpaceColRate}-\ref{eq:FreeSpacePARate})
shows that the 2D confinement brings the following effects:

\begin{itemize}
\item the 3D momentum integral becomes a 1D integral
\item the momenta involved in the integral are now typically larger than
the {}``zero-point momentum'' $\sqrt{2}\hbar/a_{0}$. Note that
$k$ is replaced by $\sqrt{2a_{0}^{-2}+p^{2}}$ in $a(k)$, and by
$2/(pa_{0}^{2})$ elsewhere.
\item there is an extra term which is real and proportional to the Hurwitz
zeta function. 
\end{itemize}
Both in 1D and 2D geometries, it is interesting to note that the photoassociation
rate now depends directly on the background scattering length $a_{\textrm{bg}}$
through $a(k)$ - see Eq. (\ref{eq:Photoassociationak}), unlike in
the free space limit. This dependence involves the ratio $a_{\textrm{bg}}/a_{0}$
of the background scattering length over the harmonic length. In many
cases, this ratio is small and can be neglected, so that the Wigner
limit (\ref{eq:Wigner1}-\ref{eq:Wigner2}) of Eqs.~(\ref{eq:2DConfinedPARate})
and (\ref{eq:1DConfinedPARate}) simplifies to:

\begin{widetext}\begin{eqnarray}
\bar{\mathcal{K}}_{\textrm{pa}}^{2D} & \!\!= & \!\!\lambda\!\!\int_{0}^{\infty}\!\frac{\frac{h}{\mu}l_{\textrm{opt}}\times e^{-\frac{E}{k_{B}T}}}{\left(\frac{E+\frac{1}{2}\hbar\Omega-E_{0}}{\gamma}\right)^{2}+\left(\frac{1}{2}+\sqrt{\pi}\frac{l_{\textrm{opt}}}{a_{0}}\right)^{2}}\frac{dE}{k_{B}T}\label{eq:2DConfinedRateWignerLimit}\end{eqnarray}
\begin{eqnarray}
\bar{\mathcal{K}}_{\textrm{pa}}^{1D} & \!\!= & \!\!\frac{\lambda^{2}}{\sqrt{\pi}}\!\!\int_{0}^{\infty}\!\frac{\frac{h}{\mu}l_{\textrm{opt}}\times e^{-\frac{E}{k_{B}T}}}{\left(\frac{E+\hbar\Omega-E_{0}}{\gamma}+\frac{l_{\textrm{opt}}}{a_{0}}\zeta(\frac{1}{2},1-\frac{E}{2\hbar\Omega})\right)^{2}+\left(\frac{1}{2}+\sqrt{\frac{2\hbar\Omega}{E}}\frac{l_{\textrm{opt}}}{a_{0}}\right)^{2}}\frac{dE}{\sqrt{k_{B}T}\sqrt{E}}\label{eq:1DConfinedRateWignerLimit}\end{eqnarray}
\end{widetext}

The result (\ref{eq:2DConfinedRateWignerLimit}) was stated in Ref.~\cite{zelevinsky2006},
omitting the factor $\lambda$ and the power-broadening term $\sqrt{\pi}l_{\textrm{opt}}/a_{0}$.
It turns out that Eq.~(\ref{eq:2DConfinedRateWignerLimit}) can be
integrated analytically:\[
\bar{\mathcal{K}}_{\textrm{pa}}^{2D}\!\!=\!\!\lambda\frac{hl_{\textrm{opt}}}{\mu}\left(\frac{\gamma}{k_{B}T}\right)^{2}F\left(\frac{E_{0}-\frac{1}{2}\hbar\Omega}{k_{B}T},\;\frac{\!\begin{array}{c}
\frac{1}{2}\end{array}\!\!+\sqrt{\pi}\frac{l_{\textrm{opt}}}{a_{0}}}{k_{B}T/\gamma}\right)\]
where $F(x,y)=\frac{1}{y}\textrm{Im}\left[e^{-(x-iy)}(\textrm{Ei}(x-iy)+i\pi)\right]$
and $\textrm{Ei}(z)=-\int_{-z}^{\infty}(e^{-t}/t)dt$ is the exponential
integral defined with a branch cut discontinuity from $-\infty$ to
0.

\subsubsection*{Quasi-2D and quasi-1D regime $\hbar\Omega\gg k_{B}T$}

At very low temperature $k_{B}T\ll\hbar\Omega$ in the 2D geometry,
we need to include the logarithmic term coming from $w$ - see Eqs.
(\ref{eq:wProperties}) - which gives the character of a true 2D collision
\cite{petrov2001}. This leads to:\begin{eqnarray}
\bar{\mathcal{K}}_{\textrm{col}}^{2D} & \!\!=\!\! & \left\langle \frac{\frac{2h}{\mu}a_{0}\times\sqrt{\pi}}{\left|\frac{a_{0}}{a(1/a_{0})}+\frac{1}{\sqrt{\pi}}\ln\left(\frac{2B/\pi}{q^{2}a_{0}^{2}}\right)+i\sqrt{\pi}\right|^{2}}\right\rangle _{\vec{q}}\label{eq:quasi2DColRate}\\
\bar{\mathcal{K}}_{\textrm{pa}}^{2D} & \!\!=\!\! & \left\langle \frac{\frac{2h}{\mu}a_{0}\times\textrm{Im}(\frac{a_{0}}{a(1/a_{0})})}{\left|\frac{a_{0}}{a(1/a_{0})}+\frac{1}{\sqrt{\pi}}\ln\left(\frac{2B/\pi}{q^{2}a_{0}^{2}}\right)+i\sqrt{\pi}\right|^{2}}\right\rangle _{\vec{q}}.\label{eq:quasi2DPARate}\end{eqnarray}
As explained in Refs.~\cite{petrov2000a,petrov2001}, the logarithmic
term leads to a {}``confinement-induced'' resonance when it cancels
the term $a_{0}/a(1/a_{0})$, which can be interpreted as the appearance
of a 2D bound state at threshold. Since we assumed that $a(k)$ already
has a resonant character - Eq. (\ref{eq:Photoassociationak}), we
may reinterpret the logarithmic term as a shift in the position of
this resonance. Note that Eq.~(\ref{eq:quasi2DColRate}) was already
considered in Ref.~\cite{wouter2003} to describe a magnetic Feshbach
resonance in a quasi-two-dimensional gas.

At very low temperature in the 1D geometry, the zeta term in Eqs.~(\ref{eq:1DConfinedColRate}-\ref{eq:1DConfinedPARate})
becomes close to $\zeta(\frac{1}{2})$: \begin{eqnarray}
\bar{\mathcal{K}}_{\textrm{col}}^{1D} & \!\!=\!\! & \left\langle \frac{\frac{2h}{\mu}a_{0}\times\frac{2}{pa_{0}}}{\left|\frac{a_{0}}{a(\sqrt{2}/a_{0})}+\zeta(\frac{1}{2})+i\frac{2}{pa_{0}}\right|^{2}}\right\rangle _{p}\label{eq:quasi1DColRate}\\
\bar{\mathcal{K}}_{\textrm{pa}}^{1D} & \!\!=\!\! & \left\langle \frac{\frac{2h}{\mu}a_{0}\times\textrm{Im}(\frac{a_{0}}{a(\sqrt{2}/a_{0})})}{\left|\frac{a_{0}}{a(\sqrt{2}/a_{0})}+\zeta(\frac{1}{2})+i\frac{2}{pa_{0}}\right|^{2}}\right\rangle _{p}.\label{eq:quasi1DPARate}\end{eqnarray}

As explained in Refs.~\cite{olshanii1998,bergeman2003a,peano2005},
the term $\zeta(\frac{1}{2})$ leads to a {}``confinement-induced''
resonance when it cancels the term $a_{0}/a(\sqrt{2}/a_{0})$. This
again can be interpreted as a shift in the position of the optical
resonance.

While the logarithmic and zeta terms in Eqs. (\ref{eq:quasi2DPARate})
and (\ref{eq:quasi1DPARate}) cause a shift in the position of the
resonance, the imaginary terms $i\sqrt{\pi}$ and $i\frac{2}{pa_{0}}$
cause a broadening of the resonance which reduces the maximal photoassociation
rate. Note that in a quasi-2D geometry, the resonance is increasingly
shifted by the logarithmic term when the temperature is decreased,
and it is broadened by a constant term $i\sqrt{\pi}$. In contrast,
in a quasi-1D geometry, the resonance is shifted by a constant term
$\zeta(\frac{1}{2})$, but is increasingly broadened by the term $i\frac{2}{pa_{0}}$
as the temperature is decreased.

The reduction of the photoassociation rate caused by this strong broadening
in 1D is related to the {}``fermionization'' of bosons \cite{girardeau1960,olshanii1998}
at low temperature, and was recently observed experimentally \cite{kinoshita2005}.
This can be understood as follows. At very low temperature the 1D
photoassociation rate Eq.~(\ref{eq:quasi1DPARate}) can be written
as\begin{equation}
\bar{\mathcal{K}}_{\textrm{pa}}^{1D}=\left\langle \vert\eta_{1D}\vert^{2}\right\rangle _{p}\times\bar{\mathcal{K}}_{\textrm{pa}}^{3D},\label{eq:FermionizationPA}\end{equation}
where $\bar{\mathcal{K}}_{\textrm{pa}}^{3D}=\frac{2h}{\mu}\times\textrm{Im}(-a)$
is the 3D photoassociation rate Eq.~(\ref{eq:FreeSpacePARate}) in
the limit $T\to0$, and $\eta_{1D}$ is the renormalisation factor
(\ref{eq:eta1D}) of the relative wave function at short separation
$r$. The average $\left\langle \vert\eta_{1D}\vert^{2}\right\rangle _{p}$
therefore corresponds to the pair correlation function $g_{2}$ (of
the two-body problem) at short separation. Fermionization occurs when
this quantity becomes small, \emph{ie} the probability of finding
atoms at short separation is strongly decreased. As photoassociation
typically occurs at these short separations, the photoassociation
rate (\ref{eq:FermionizationPA}) is strongly reduced. For the Boltzmann
distribution $\mathcal{P}(p)$ that we have assumed so far, $\left\langle \vert\eta_{1D}\vert^{2}\right\rangle _{p}$
actually vanishes at $T=0$. A more proper calculation using the relative
momentum distribution \cite{olshanii2003} in the fermionization regime
leads to $\left\langle \vert\eta_{1D}\vert^{2}\right\rangle _{p}=\frac{4}{3}\pi^{2}\left(\frac{2\vert a\vert}{n_{1D}a_{0}^{2}}\right)^{-2}$,
where $n_{1D}$ is the 1D density of the gas, in accordance with the
calculation of the pair correlation function $g_{2}$ in the many-body
framework of Ref.~\cite{gangardt2003} (note however that $a$ denotes
the modified scattering length (\ref{eq:ModifiedScatteringLength}),
instead of the background scattering length).

\subsection{Position of the resonance\label{sub:Position-of-the-resonance}}

In a typical experiment, the collision or photoassociation rates are
measured as a function of the laser frequency. In a lattice, the frequency
for which the laser is resonant (\emph{ie} which gives a peak in the
rate) may be shifted from the expected free space resonant frequency.
The lattice eliminates the recoil shift but also introduces new shifts.
In our equations, the free-space resonance is obtained for $E_{0}=0$,
therefore we define the lattice shift $\Delta E$ as the value of
$E_{0}$ which maximizes (\ref{eq:2DPhotoassociationRate}) or (\ref{eq:1DPhotoassociationRate}).

There are several contributions to the shift. The first one is the
zero-point energy, which can be easily understood as follows. The
relative motion of the ground-state atoms is influenced by the confinement,
and therefore has the zero-point energy, even for $T=0$. On the other
hand, the relative motion in the excited molecular bound state is
not affected by the confinement (only the centre-of-mass is). As a
result, the transition is shifted by the zero-point energy.

The thermal average in (\ref{eq:2DPhotoassociationRate}) and (\ref{eq:1DPhotoassociationRate})
also shifts the line by a quantity of order $k_{B}T$. Due to the
dimensionality of the integral involved in the averaging, this thermal
shift will differ from the usual 3D thermal shift.

In the 2D confinement-dominated regime, we find the following shift:\begin{equation}
\Delta E_{\textrm{CD}}=\frac{1}{2}\hbar\Omega+k_{B}T\times Z\left(\frac{\!\begin{array}{c}
\frac{1}{2}\end{array}\!\!+\sqrt{\pi}\frac{l_{\textrm{opt}}}{a_{0}}}{k_{B}T/\gamma}\right).\label{eq:shift1}\end{equation}
The first term is the zero-point energy shift and the second term
is the thermal shift, where $Z(y)\approx1-(2/y)^{2}$ is the solution
of $F(Z,y)=\frac{1}{Z^{2}+y^{2}}$, and $F$ is the function defined
previously.

We mentioned that the background scattering length $a_{\textrm{bg}}$
affects the photoassociation rate in a lattice, and in particular
it may shift the line if it is large enough (note that this is an
intrinsic molecular shift, not a density-dependent shift due to interaction
with other atoms \cite{ido2005}), however we find this shift to be
very small in most cases.

Finally, at high intensity and low temperature, the line may be shifted
by the logarithmic term in a 2D geometry, or by the zeta term in a
1D geometry. The 2D logarithmic shift is potentially the most easily
observable experimentally, because the line is not as broadened as
in the 1D case. We find that the following formula gives a good approximation
of the shift in the quasi-2D regime at high intensity:\begin{equation}
\Delta E_{2D}\approx\frac{1}{2}\hbar\Omega+k_{B}T+\frac{l_{\textrm{opt}}}{a_{0}}\frac{\gamma}{\sqrt{\pi}}\left(\frac{1}{2}+\ln\left(\frac{B\hbar\Omega}{\pi k_{B}T}\right)\right)\label{eq:shift2}\end{equation}

Interestingly, the {}``logarithmic shift'' is proportional to the
optical length $l_{\textrm{opt}}$. In free space, the optical length
is usually determined by measuring the rate (\ref{eq:FreeSpaceRateWignerLimit})
which is proportional to it. However the rate can typically have a
factor of 2 uncertainty as its determination relies on the knowledge
of the density of atoms \cite{michelson2005,zelevinsky2006}. Here
we propose a determination of $l_{\textrm{opt}}$ which does not require
the knowledge of the density and is is simply based on a spectroscopic
measurement: by measuring the position of the line for different intensities,
one should observe a shift linear with the intensity $I$, and deduce
the strength of the resonance $l_{\textrm{opt}}/I$. The major constraint
of this method is to be able to reach and measure sufficiently low
temperatures $\ll\hbar\Omega$. One should also take into account
any possible light shift \cite{ciurylo2006} (also proportional to
the intensity) by measuring it at higher temperatures.

\subsection{Line shape and strength}

The shape and apparent strength of the line are also modified by the
lattice. In the confinement-dominated regime, we saw that the dimensionality
of the thermal distribution is reduced. As a result, the corresponding
density of states ($\sim1$ in 2D, and $\sim E^{-1/2}$ in 1D) gives
more weight to small energies - compare Eqs.~(\ref{eq:2DConfinedRateWignerLimit})
and (\ref{eq:1DConfinedRateWignerLimit}) with Eq.~(\ref{eq:FreeSpaceRateWignerLimit}).
The line is therefore expected to be narrower, especially in the 1D
case. The precise strength of the line depends on the density and
the thermal averaging, but we can get some qualitative insight by
considering the on-resonance rate in the confinement-dominated regime,
disregarding the thermal average.

In most cases, species have a small scattering length $a_{\textrm{bg}}\ll a_{0}$.
In this limit, the on-resonance rates are essentially:\begin{eqnarray}
\bar{\mathcal{K}}_{\textrm{pa, res}}^{2D} & \sim & \frac{4h}{\mu}l_{\textrm{opt}}\frac{1}{\left(1+2\sqrt{\pi}\frac{l_{\textrm{opt}}}{a_{0}}\right)^{2}}\label{eq:PowerBroadening2D}\\
\bar{\mathcal{K}}_{\textrm{pa, res}}^{1D} & \sim & \frac{4h}{\mu}l_{\textrm{opt}}\frac{1}{\left(1+4\frac{l_{\textrm{opt}}}{pa_{0}^{2}}\right)^{2}}\label{eq:PowerBroadening1D}\end{eqnarray}

One can see that the power broadening term $kl_{\textrm{opt}}$ in
Eq.~(\ref{eq:FreeSpaceRateWignerLimit}) is enhanced because the
momentum $\hbar k$ is now bounded from below by the zero-point momentum
$\sim\hbar/a_{0}$. It is even much larger at small temperature in
the quasi-1D regime, because of the fermionization effect discussed
earlier. It is therefore easier in a lattice to saturate the line
with lower laser intensities, even at very small temperature.

On the other hand, certain species may have a large scattering length
$a_{\textrm{bg}}\sim a_{0}$. In this case, the rate explicitly depends
on the scattering length. For small laser intenstites ($l_{\textrm{opt}}\ll a_{\textrm{bg}},\; a_{0}$),
the on-resonance rates are essentially:\begin{eqnarray}
\bar{\mathcal{K}}_{\textrm{pa, res}}^{2D} & \sim & \frac{4h}{\mu}l_{\textrm{opt}}\frac{1+(a_{\textrm{bg}}/a_{0})^{2}}{1+\pi(a_{\textrm{bg}}/a_{0})^{2}}\label{eq:BackgroundDependence2D}\\
\bar{\mathcal{K}}_{\textrm{pa, res}}^{1D} & \sim & \frac{4h}{\mu}l_{\textrm{opt}}\frac{1+2(a_{\textrm{bg}}/a_{0})^{2}}{\left(1+\frac{a_{\textrm{bg}}}{a_{0}}\zeta(\frac{1}{2})\right)^{2}+\left(\frac{2a_{\textrm{bg}}}{pa_{0}^{2}}\right)^{2}}\label{eq:BackgroundDependence1D}\end{eqnarray}

We give an illustration of this dependence on $a_{\textrm{bg}}$ in
the next Section in the case of Strontium 86.

In principle, this new dependence on the background scattering length
could be used as a way to improve its determination when it is large.
This would bring complementary information to the usual node method
\cite{michelson2005} which proves to be difficult for large scattering
lengths as it requires a very precise knowledge of the long-range
part of the potential. However,  the typical uncertainties associated
to measured rates in current experiments may limit the interest of
this idea.

\section{Alkaline-Earth photoassociation in a 1D or 2D lattice\label{sec:AlkalineEarth}}

\begin{figure*}[t]
\includegraphics[%
  scale=0.35]{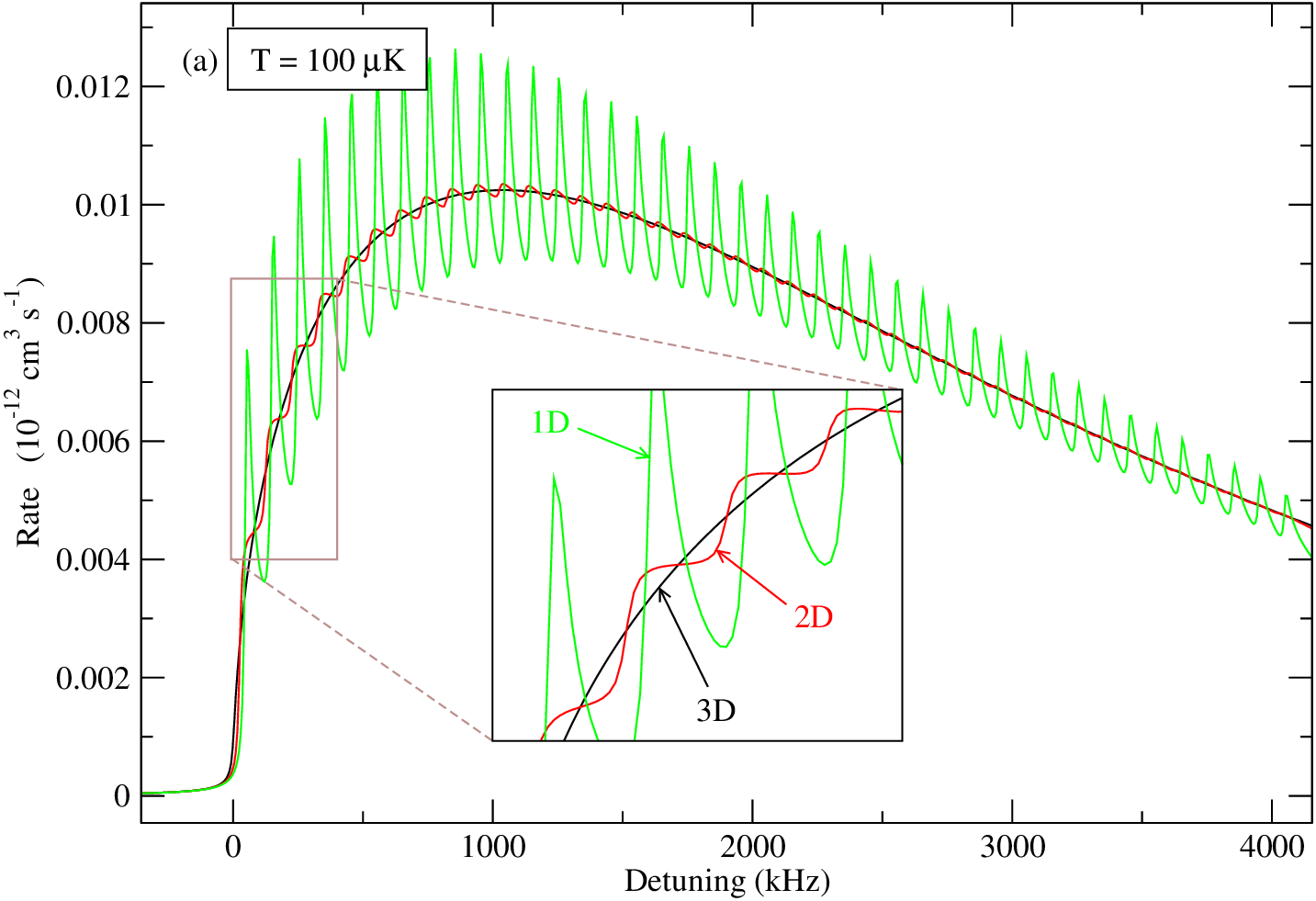}\includegraphics[%
  scale=0.35]{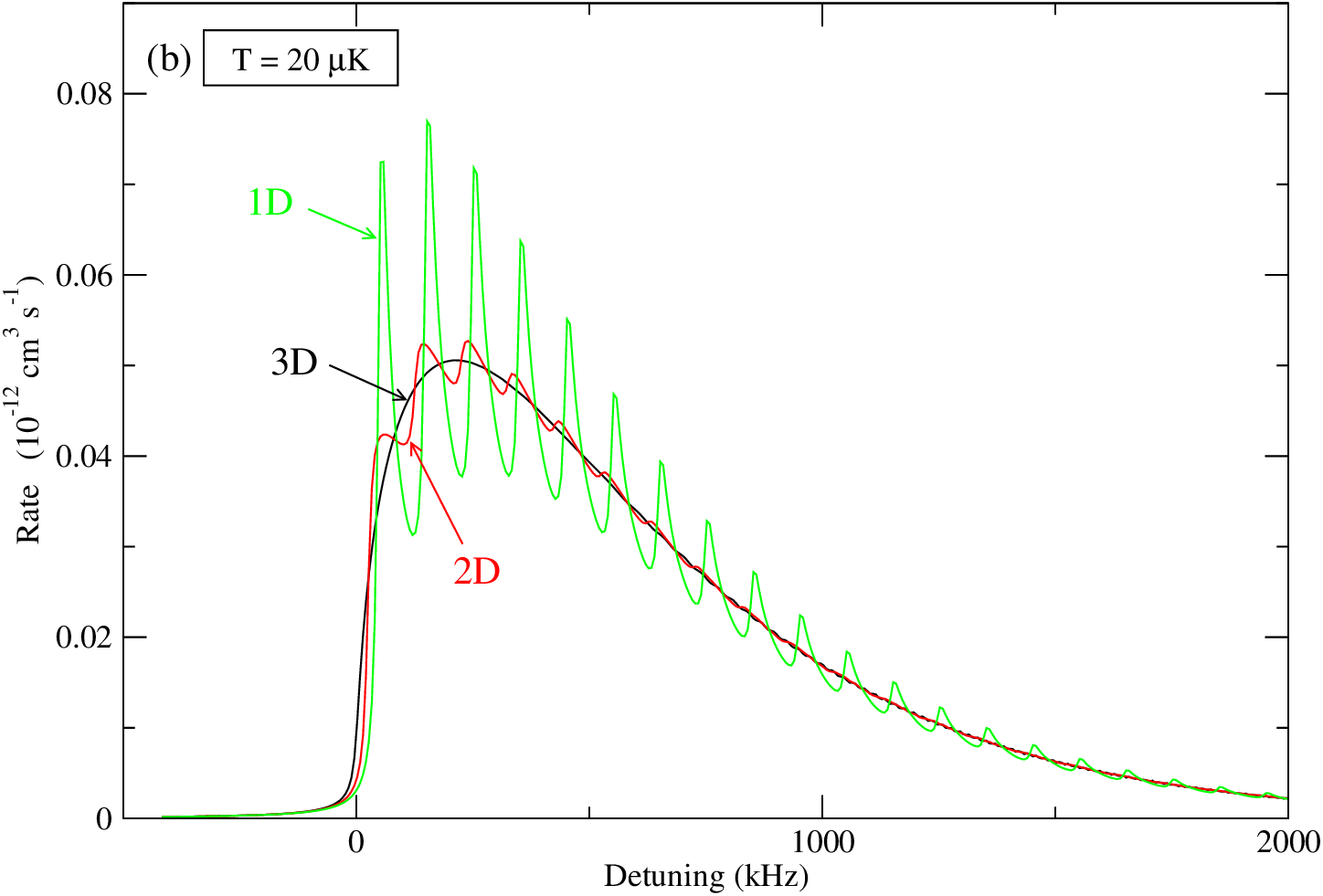}\\
\includegraphics[%
  clip,
  scale=0.35]{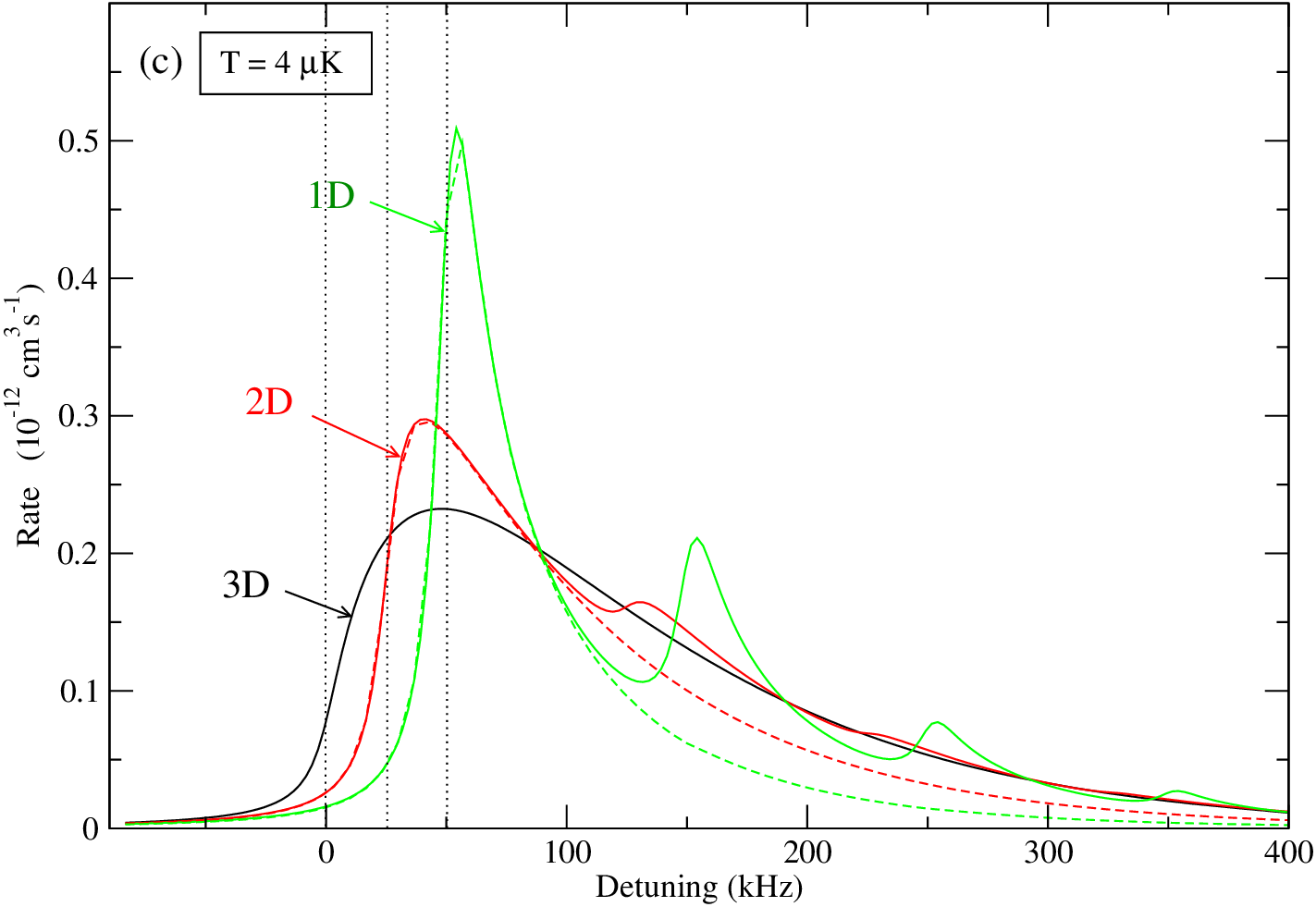}\includegraphics[%
  clip,
  scale=0.35]{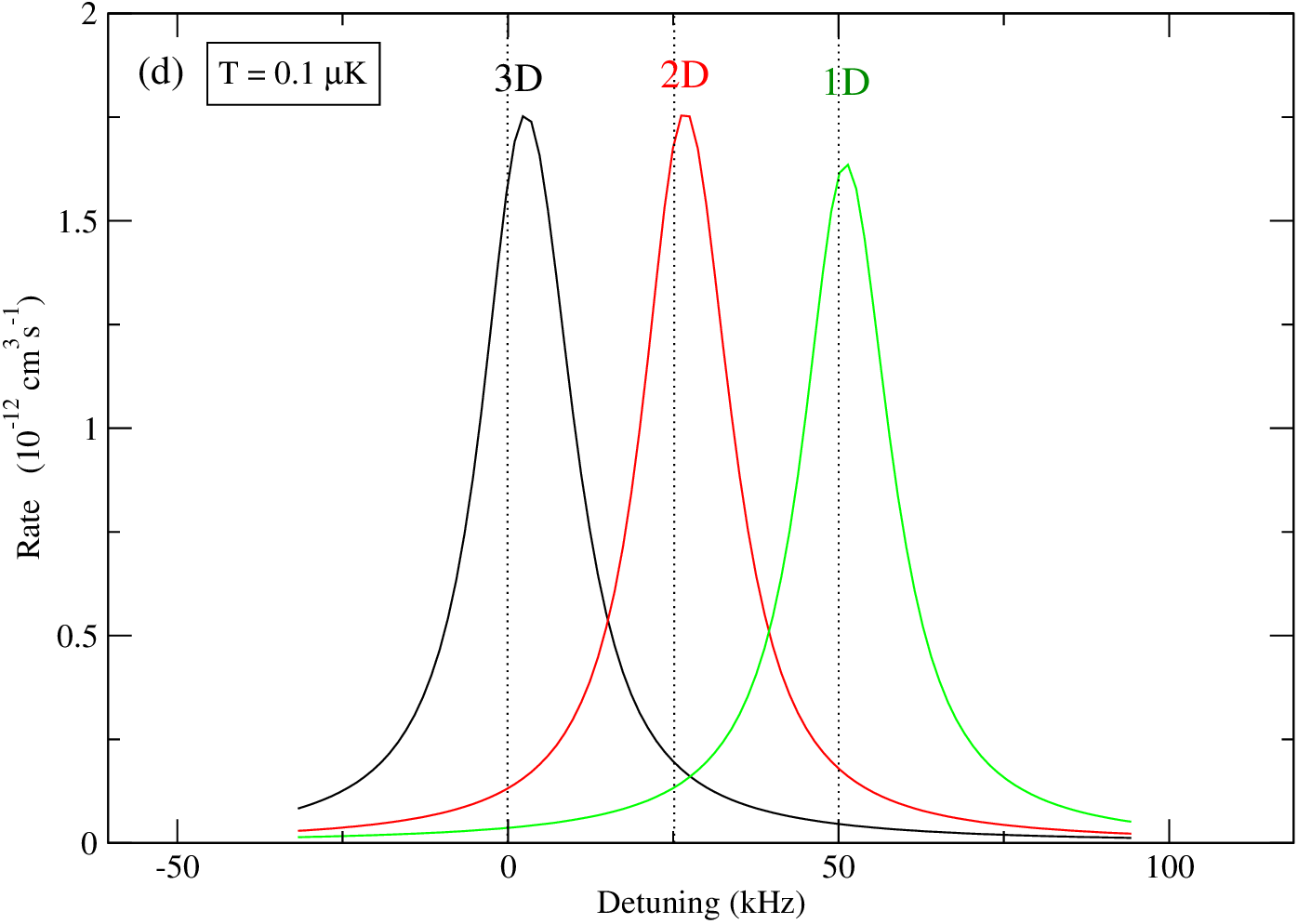}

\caption{\label{cap:Strontium88Low}Photoassociation rate of Strontium 88
at low intensity ($l_{\textrm{opt}}=1$ Bohr $\approx$ 0.0529 nm)
as a function of laser detuning, comparing free-space (3D), 1D-confinement
(2D geometry) and 2D-confinement (1D geometry), with a trapping frequency
$\Omega=2\pi\times50$ kHz. Black curves are the free-space 3D prediction
- Eq.~(\ref{eq:FreeSpacePARate}). Red curves are the effective 3D
rate in a 2D geometry - Eq.~(\ref{eq:2DPhotoassociationRate}). Green
curves are the effective 3D rate in a 1D geometry - Eq.~(\ref{eq:1DPhotoassociationRate}).
Panel (a)~: $T=100\,\mu$K (free-space regime), Panel (b)~: $T=20\,\mu$K;
Panel (c)~: $T=$ 4 $\mu$K (confinement-dominated regime), Panel
(d)~: $T=$ 0.1 $\mu$K (Quasi-2D or -1D regime). In the confinement-dominated
regime, the approximate line shapes calculated from Eqs. (\ref{eq:2DConfinedPARate}-\ref{eq:1DConfinedPARate})
(considering only the trap ground-state contribution) are represented
by dashed curves.}
\end{figure*}

We illustrate the previous theory by considering alkaline-earth intercombination
photoassociation in a tight optical lattice. An important feature
of the intercombination line is the low spontaneous emission from
the photoassociated state. While the natural width for allowed transitions
is typically tens of MHz, the width of the intercombination lines
is typically tens of kHz. This is on the same order of magnitude as
the typical trapping frequency at the centre of each lattice cell.
As a result, we expect a clear manifestation of the effects described
in the previous section for alkaline-earth intercombination transitions.
On the other hand, they should hardly be observable with allowed transitions,
as they would be hidden by the large natural width.

\begin{figure}
\includegraphics[%
  scale=0.35]{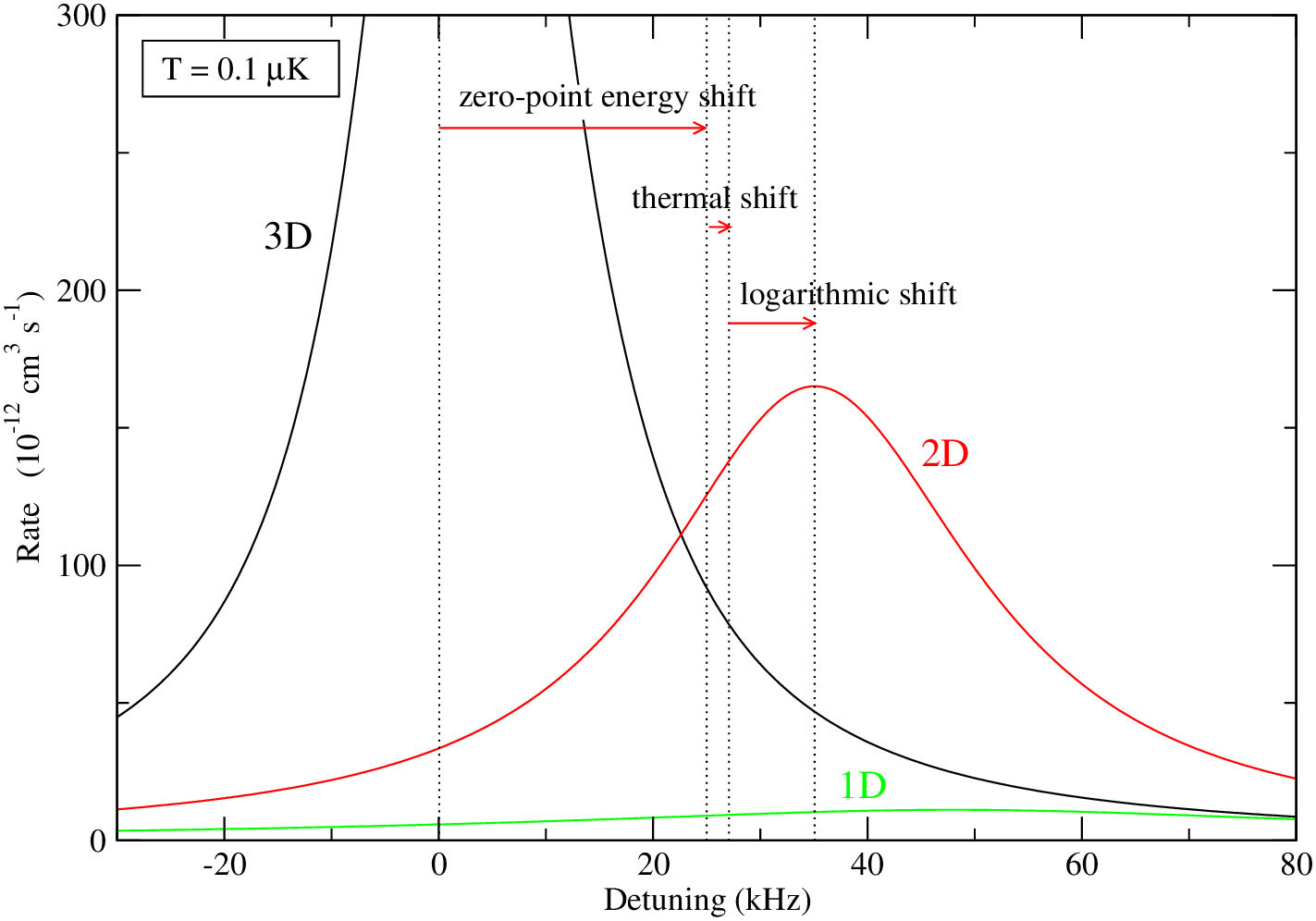}

\caption{\label{cap:Strontium88High}Same as Fig. \ref{cap:Strontium88Low}d
for higher intensity or stronger resonance ($l_{\textrm{opt}}=500$
Bohr $\approx$ 26.6 nm). The 2D and 1D lines are now saturated (especially
the 1D line), in accordance with Eqs.~(\ref{eq:PowerBroadening2D})
and (\ref{eq:PowerBroadening1D}), while the free-space line is not.
The 2D line is also shifted according to Eq.~(\ref{eq:shift2}).
We indicated the three contributions to the shift by horizontal arrows.}
\end{figure}

\begin{figure}
\includegraphics[%
  bb=23bp 42bp 706bp 542bp,
  scale=0.35]{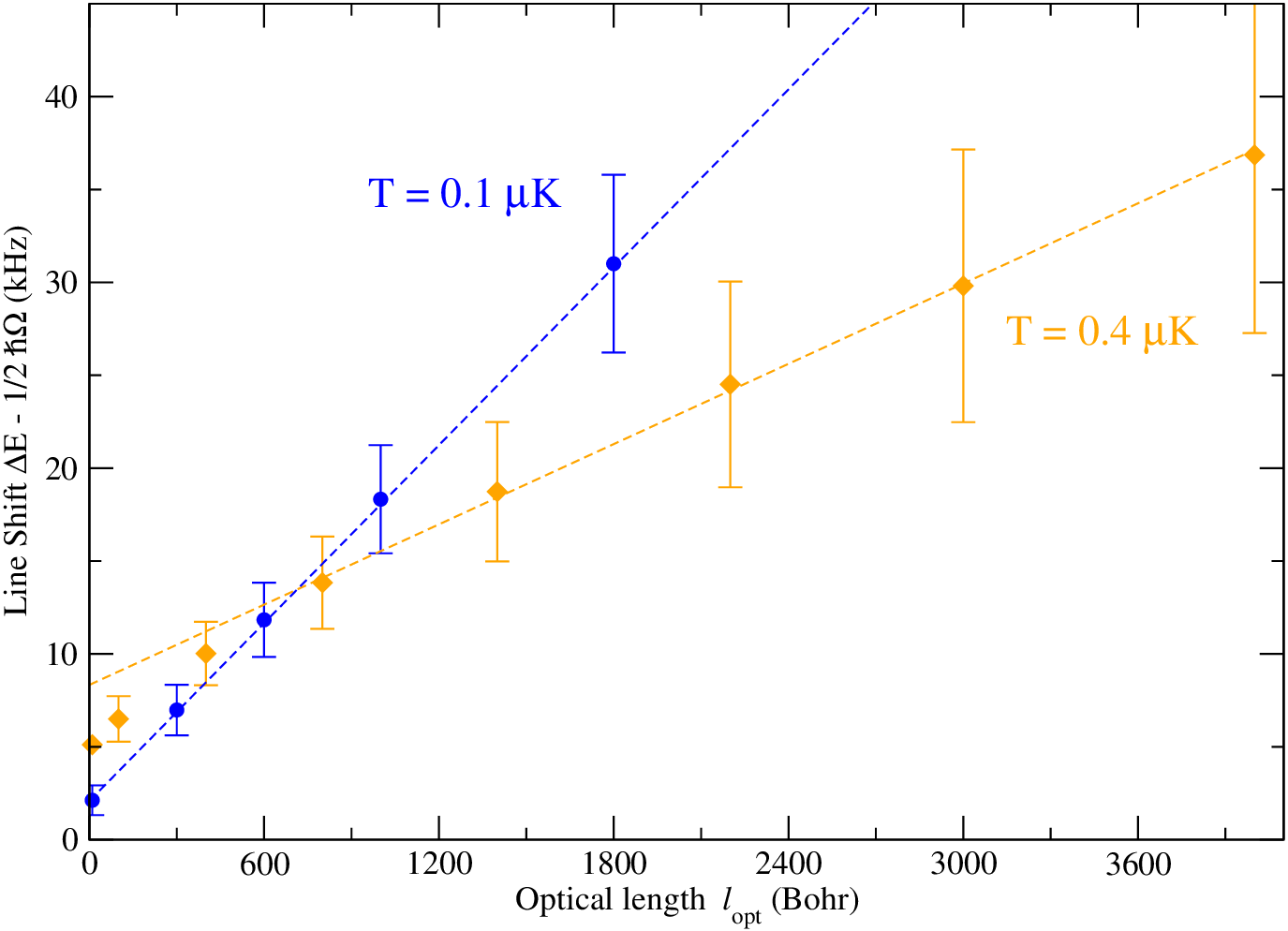}

\caption{\label{cap:Strontium88shift}Shift $\Delta E$ of the photoassociation
line in a 2D geometry with trap frequency $\Omega=2\pi\times50$ kHz,
as a function of optical length (proportional to photoassociation
laser intensity). For convenience, we substracted the zero-point energy
shift $\frac{1}{2}\hbar\Omega$ which is constant. The orange diamonds
correspond to a temperature $T=0.4\,\mu$K, and the blue dots corresponds
to a temperature $T=0.1\,\mu$K. These points are obtained by calculating
the line shape using Eq.~(\ref{eq:quasi2DPARate}) and finding the
shift numerically. For each point, we represented a vertical line
whose length corresponds to one sixth of the photoassociation line
width; this gives an indication of how accurately the shift can be
determined experimentally. The dashed lines are obtained using the
approximate formula Eq.~(\ref{eq:shift2}).}
\end{figure}

\begin{figure}
\hfill{}\begin{tabular}{c}
\includegraphics[%
  bb=23bp 42bp 706bp 542bp,
  scale=0.32]{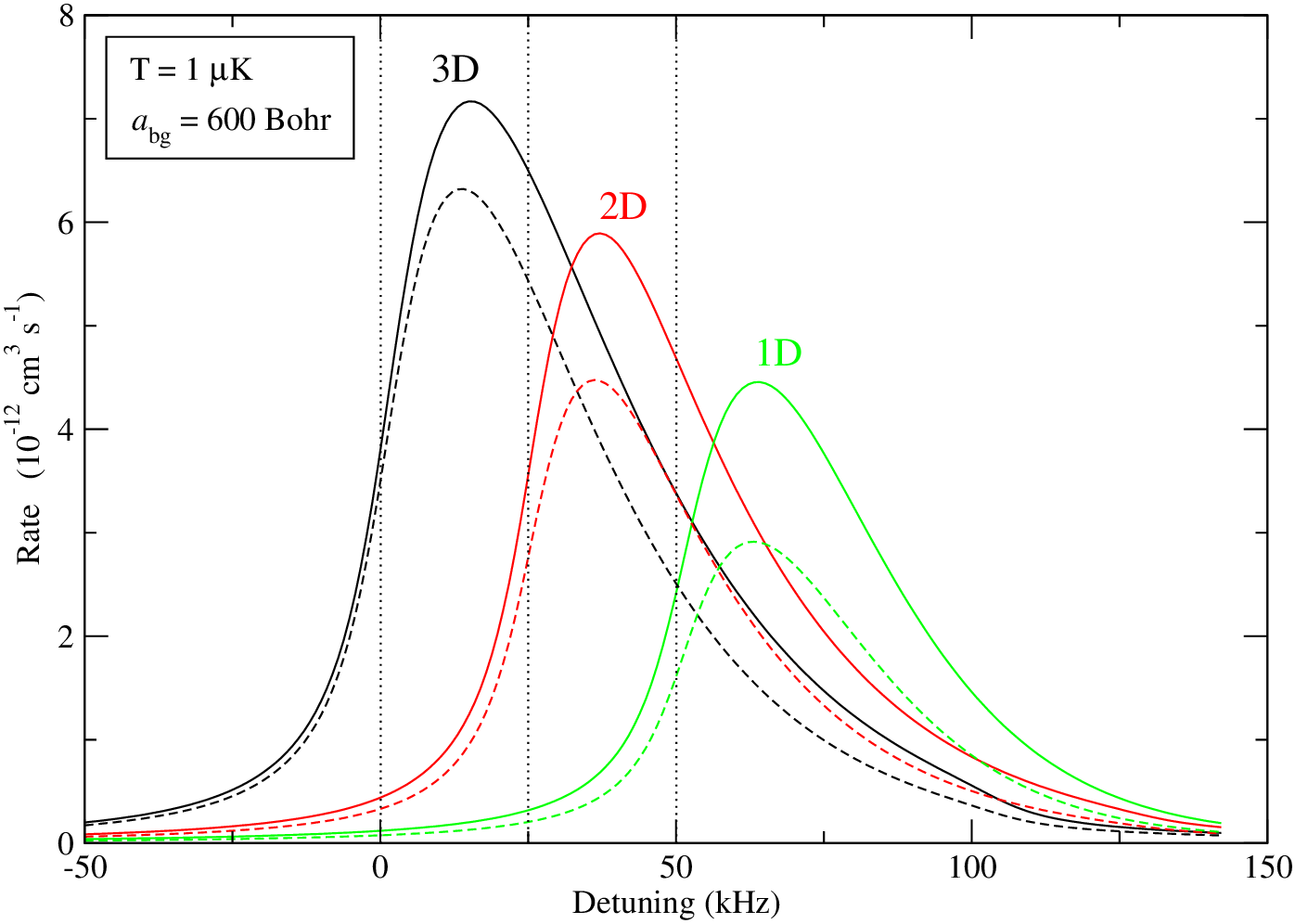}\tabularnewline
\includegraphics[%
  clip,
  scale=0.32]{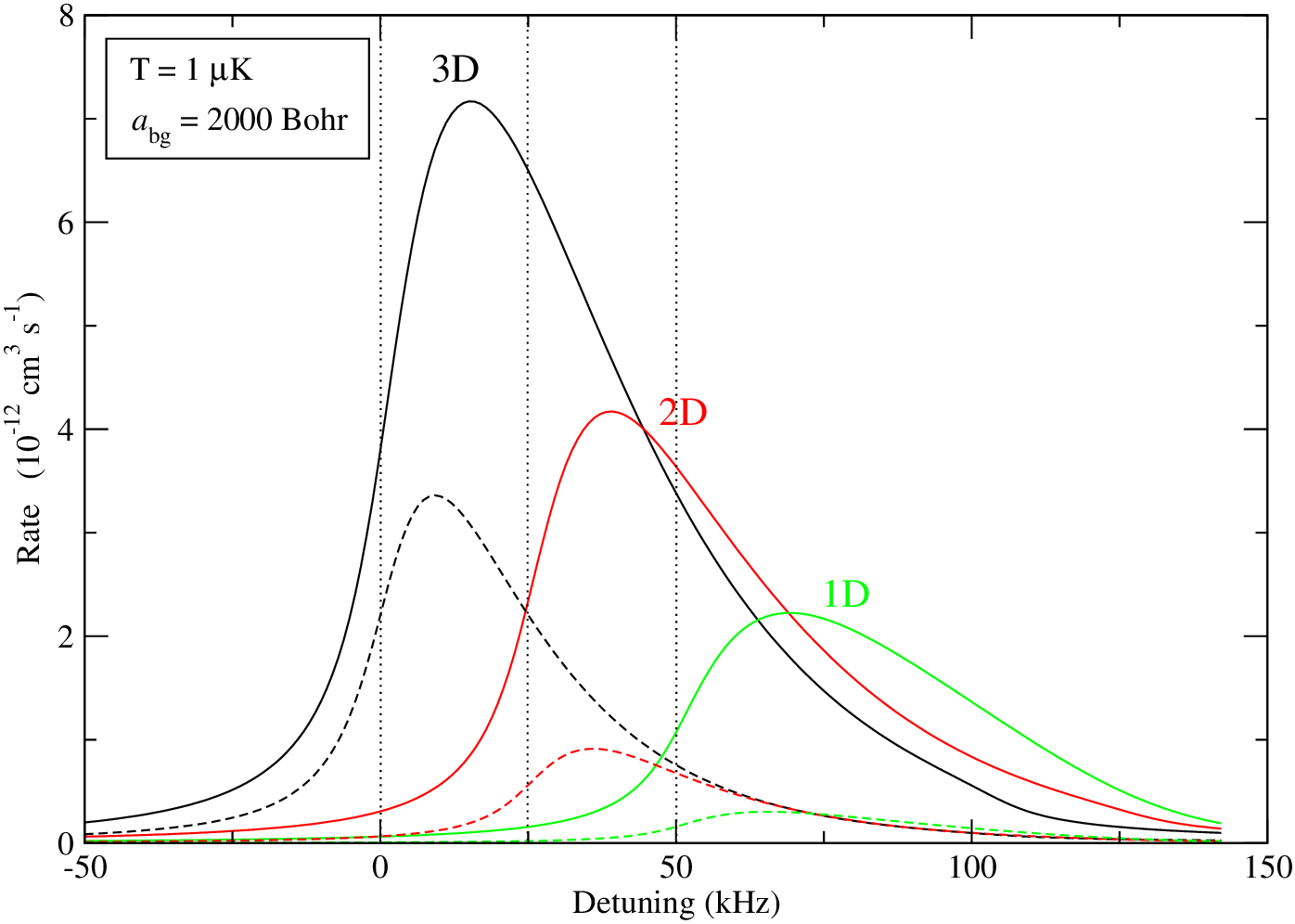}\tabularnewline
\end{tabular}\hfill{}

\caption{\label{cap:Strontium86}Photoassociation rates for Strontium 86 ($\Omega=2\pi\times50$
kHz, $l_{\textrm{opt}}=10$ Bohr $\approx$ 0.0529 nm, $T=1\,\mu$K;
conditions similar to the experiment of Ref.~\cite{zelevinsky2006}).
Black curves are the free-space 3D rate prediction, red and green
curves are the effective 3D rate from the 2D and 1D theories, using
the effective scattering length (\ref{eq:Photoassociationak}); the
dashed curves are obtained using the more usual expression (\ref{eq:ModifiedScatteringLength})
for the scattering length. Panel (a): assuming $a_{\textrm{bg}}=600$
Bohr $\approx$ 31.75 nm. Panel (b): assuming $a_{\textrm{bg}}=2000$
Bohr $\approx$ 105.8 nm. Note that the free-space rate does not depend
on the background scattering length $a_{\textrm{bg}}$.}
\end{figure}

Figure \ref{cap:Strontium88Low} shows the effective photoassociation
rate at different temperatures for a typical intercombination line
of Strontium 88 in free space, 1D lattice, and 2D lattice. The scattering
length of Strontium is taken to be 10 Bohr \cite{michelson2005} (1
Bohr $\approx$ 0.0529177 nm) and the spontaneous emission term $\gamma=15$
kHz \cite{zelevinsky2006}. The trapping frequency $\Omega$ is 2$\pi\times$50
kHz, and the optical length is set to 1 Bohr. The temperatures correspond
to the three regimes discussed in Section \ref{sub:Comparison-of-2D-1D}:
free space ($T=100\,\mu$K and $T=20\,\mu$K), confinement-dominated
($T=4\,\mu$K ), and 2D regime ($T=0.1\,\mu$K ). 

For $T=100\,\mu$K and $T=20\,\mu$K (Fig.~\ref{cap:Strontium88Low}a
and \ref{cap:Strontium88Low}b), one might expect the line shape to
be very close to the free-space theory prediction. This is indeed
the case in a 2D geometry. However, higher temperatures are needed
to reach the free-space regime in a 1D geometry. For the temperatures
considered here, we observe a lot of distinct spikes in the 1D photoassociation
line. These spikes correspond to contributions from different trap
states and are therefore separated by twice the frequency $\Omega$
of the trap - see Eq.~(\ref{eq:Energy1D}) with $m=0$. There are
similar contributions in the 2D case, but they tend to overlap, due
to their width. In contrast, the spikes are very narrow in the 1D
case, because the one-dimensional density of states scales like $1/\sqrt{E}$
(see, for instance, Eq.(\ref{eq:1DConfinedRateWignerLimit})), which
enhances low-energy collisions. As a result each spike has a width
close to the natural width of the resonance and is well separated
from the others, even at relatively high temperatures (tens of $\mu$K)
which are more than one order of magnitude larger than the natural
width. This feature could be exploited to measure the position of
the line at nearly the natural width precision, while operating at
temperatures much larger than this width. Note that we have assumed
ideal harmonic confinement, which is why the spikes are regularly
spaced and well separated. Experimentally, only the first spikes should
be visible, due to the anharmonicity of the lattice potential.

In the confinement-dominated regime (Fig.~\ref{cap:Strontium88Low}c),
the line shape is dominated by the first spike (contribution from
the trap ground state) and the line is clearly shifted due to the
zero-point energy. In the 2D case, the density of states leads to
a line which is more symmetric and narrower than in 3D, resulting
in an increase of the photoassociation rate on resonance with respect
to the free-space rate. This situation corresponds roughly to the
experimental conditions of Ref.~\cite{zelevinsky2006}. However,
the experimental conditions were such that the 2D and 3D theory happened
to predict very similar rates, and extra broadening mechanisms made
it difficult to observe an unambiguous 2D line shape as in Fig.~\ref{cap:Strontium88Low}c.
In the 1D case, the thermal distribution leads to an even narrower
line and higher rate, as small collision energies are enhanced by
the density of states.

In the quasi-2D or quasi-1D regime (Fig.~\ref{cap:Strontium88Low}d),
the line is almost unbroadened by temperature and recovers its natural
Lorentzian shape in all theories. The temperature is so low that there
is almost one collision energy in the system: 0 in the 3D theory,
$\frac{1}{2}\hbar\Omega$ in the 2D theory, and $\hbar\Omega$ in
the 1D theory.

In Fig.~\ref{cap:Strontium88High}, we consider a higher optical
length ($l_{\textrm{opt}}=500$ Bohr), which can be reasonably attained
using the intercombination lines close to thresold, or by increasing
the photoassociation laser intensity. The line is now power-broadened,
as predicted by Eqs.~(\ref{eq:PowerBroadening2D}-\ref{eq:PowerBroadening1D}),
while the free-space line is not. As explained before, the broadening
is due to the zero-point momentum in the 2D case, and to fermionization
in the 1D case. In the quasi-2D regime, the line is also shifted by
the logarithmic term, according to Eq.~(\ref{eq:shift2}). As suggested
in Section \ref{sub:Position-of-the-resonance}, this shift could
be used as a way to measure the optical length. However, the line
is progressively power-broadened as it is shifted. In Fig.~\ref{cap:Strontium88shift},
we plot the shift and width as a function of the optical length. Although
the experiment may not be easy, this graph suggests that the shift
should be observable.

We now illustrate the theory with the intercombination lines in Strontium
86. The background scattering length of this species has been found
to be quite large, between 610 and 2300 Bohr \cite{michelson2005}.
As a result, the photoassociation line in the lattice is notably modified
by the scattering length. In Fig.~\ref{cap:Strontium86}, we plotted
the photoassociation rate for $a_{bg}=600$ Bohr and $a_{bg}=$ 2000
Bohr. The effect of the scattering length is clearly noticeable, as
predicted qualitatively by Eqs.~(\ref{eq:BackgroundDependence2D}-\ref{eq:BackgroundDependence1D}).
As noted previousy, measurements of the rate for different lattice
wave lengths might be able to give a better constraint on the scattering
length.

We also plotted on Fig.~\ref{cap:Strontium86} the effective rate
obtained if one replaces the effective scattering length (\ref{eq:Photoassociationak})
by the more usual resonant scattering length (\ref{eq:ModifiedScatteringLength}).
The latter appears to be insufficient to predict the correct rate
for large scattering length, showing that in this particular case,
one cannot simply extrapolate the original work of Refs.~\cite{petrov2001,olshanii1998}
by replacing the scattering length by the resonant scattering length
(\ref{eq:ModifiedScatteringLength}).

\section{Conclusion\label{sec:Conclusion}}

We investigated how photoassociation is affected by the confinement
of a 1D or 2D optical lattice. We showed that effects are expected
to be observable in the case of narrow resonances, like the intercombination
lines found in alkaline-earth species. The main effects of confinement
are: shift of the resonance (most importantly by the zero-point energy
of the lattice), narrower lines, increase of power-broadening at low
temperature, and direct sensitivity to a large value of the scattering
length of the atoms. More specifically, a 2D confinement could significantly
increase the spectroscopic resolution, making it possible to observe
the natural line width at relatively high temperatures. At low temperature,
a 1D confinement induces a shift of the line which is proportional
to the intensity of the photoassociation laser. This shift could be
used to determine the strength of a resonance, without having to measure
the density of the system.

\begin{acknowledgments}
We thank T. Zelevinsky and K. Enomoto for their helpful comments.
This work has been partially supported by the U.S. Office of Naval
Research.
\end{acknowledgments}
\bibliographystyle{apsrev}
\bibliography{/home/pascal/Redaction/biblio,/home/pascal/Redaction/biblio_extra,/Users/pascal/Documents/Travail/Redaction/biblio,/Users/pascal/Documents/Travail/Redaction/biblio_extra}

\begin{thebibliography}{39}
\expandafter\ifx\csname natexlab\endcsname\relax\def\natexlab#1{#1}\fi
\expandafter\ifx\csname bibnamefont\endcsname\relax
  \def\bibnamefont#1{#1}\fi
\expandafter\ifx\csname bibfnamefont\endcsname\relax
  \def\bibfnamefont#1{#1}\fi
\expandafter\ifx\csname citenamefont\endcsname\relax
  \def\citenamefont#1{#1}\fi
\expandafter\ifx\csname url\endcsname\relax
  \def\url#1{\texttt{#1}}\fi
\expandafter\ifx\csname urlprefix\endcsname\relax\def\urlprefix{URL }\fi
\providecommand{\bibinfo}[2]{#2}
\providecommand{\eprint}[2][]{\url{#2}}

\bibitem[{\citenamefont{Katori et~al.}(2003)\citenamefont{Katori, Takamoto,
  Pal'chikov, and Ovsiannikov}}]{katori2003}
\bibinfo{author}{\bibfnamefont{H.}~\bibnamefont{Katori}},
  \bibinfo{author}{\bibfnamefont{M.}~\bibnamefont{Takamoto}},
  \bibinfo{author}{\bibfnamefont{V.~G.} \bibnamefont{Pal'chikov}},
  \bibnamefont{and} \bibinfo{author}{\bibfnamefont{V.~D.}
  \bibnamefont{Ovsiannikov}}, \bibinfo{journal}{Phys. Rev. Lett.}
  \textbf{\bibinfo{volume}{91}}, \bibinfo{pages}{173005}
  (\bibinfo{year}{2003}).

\bibitem[{\citenamefont{Ido et~al.}(2005)\citenamefont{Ido, Loftus, Boyd,
  Ludlow, Holman, and Ye}}]{ido2005}
\bibinfo{author}{\bibfnamefont{T.}~\bibnamefont{Ido}},
  \bibinfo{author}{\bibfnamefont{T.~H.} \bibnamefont{Loftus}},
  \bibinfo{author}{\bibfnamefont{M.~M.} \bibnamefont{Boyd}},
  \bibinfo{author}{\bibfnamefont{A.~D.} \bibnamefont{Ludlow}},
  \bibinfo{author}{\bibfnamefont{K.~W.} \bibnamefont{Holman}},
  \bibnamefont{and} \bibinfo{author}{\bibfnamefont{J.}~\bibnamefont{Ye}},
  \bibinfo{journal}{Phys. Rev. Lett.} \textbf{\bibinfo{volume}{94}},
  \bibinfo{pages}{153001} (\bibinfo{year}{2005}).

\bibitem[{\citenamefont{Ludlow et~al.}(2006)\citenamefont{Ludlow, Boyd,
  Zelevinsky, Foreman, Blatt, Notcutt, Ido, and Ye}}]{ludlow2006}
\bibinfo{author}{\bibfnamefont{A.~D.} \bibnamefont{Ludlow}},
  \bibinfo{author}{\bibfnamefont{M.~M.} \bibnamefont{Boyd}},
  \bibinfo{author}{\bibfnamefont{T.}~\bibnamefont{Zelevinsky}},
  \bibinfo{author}{\bibfnamefont{S.~M.} \bibnamefont{Foreman}},
  \bibinfo{author}{\bibfnamefont{S.}~\bibnamefont{Blatt}},
  \bibinfo{author}{\bibfnamefont{M.}~\bibnamefont{Notcutt}},
  \bibinfo{author}{\bibfnamefont{T.}~\bibnamefont{Ido}}, \bibnamefont{and}
  \bibinfo{author}{\bibfnamefont{J.}~\bibnamefont{Ye}}, \bibinfo{journal}{Phys.
  Rev. Lett.} \textbf{\bibinfo{volume}{96}}, \bibinfo{pages}{033003}
  (\bibinfo{year}{2006}).

\bibitem[{\citenamefont{Takasu et~al.}(2003)\citenamefont{Takasu, Maki, Komori,
  Takano, Honda, Kumakura, Yabuzaki, and Takahashi}}]{takasu2003}
\bibinfo{author}{\bibfnamefont{Y.}~\bibnamefont{Takasu}},
  \bibinfo{author}{\bibfnamefont{K.}~\bibnamefont{Maki}},
  \bibinfo{author}{\bibfnamefont{K.}~\bibnamefont{Komori}},
  \bibinfo{author}{\bibfnamefont{T.}~\bibnamefont{Takano}},
  \bibinfo{author}{\bibfnamefont{K.}~\bibnamefont{Honda}},
  \bibinfo{author}{\bibfnamefont{M.}~\bibnamefont{Kumakura}},
  \bibinfo{author}{\bibfnamefont{T.}~\bibnamefont{Yabuzaki}}, \bibnamefont{and}
  \bibinfo{author}{\bibfnamefont{Y.}~\bibnamefont{Takahashi}},
  \bibinfo{journal}{Phys. Rev. Lett.} \textbf{\bibinfo{volume}{91}},
  \bibinfo{pages}{040404} (\bibinfo{year}{2003}).

\bibitem[{\citenamefont{Nagel et~al.}(2005)\citenamefont{Nagel, Mickelson,
  Saenz, Martinez, Chen, Killian, Pellegrini, and Côt\'{e}}}]{nagel2005}
\bibinfo{author}{\bibfnamefont{S.~B.} \bibnamefont{Nagel}},
  \bibinfo{author}{\bibfnamefont{P.~G.} \bibnamefont{Mickelson}},
  \bibinfo{author}{\bibfnamefont{A.~D.} \bibnamefont{Saenz}},
  \bibinfo{author}{\bibfnamefont{Y.~N.} \bibnamefont{Martinez}},
  \bibinfo{author}{\bibfnamefont{Y.~C.} \bibnamefont{Chen}},
  \bibinfo{author}{\bibfnamefont{T.~C.} \bibnamefont{Killian}},
  \bibinfo{author}{\bibfnamefont{P.}~\bibnamefont{Pellegrini}},
  \bibnamefont{and} \bibinfo{author}{\bibfnamefont{R.}~\bibnamefont{Côt\'{e}}},
  \bibinfo{journal}{Phys. Rev. Lett.} \textbf{\bibinfo{volume}{94}},
  \bibinfo{pages}{083004} (\bibinfo{year}{2005}).

\bibitem[{\citenamefont{Takasu et~al.}(2004)\citenamefont{Takasu, Komori,
  Honda, Kumakura, Yabuzaki, and Takahashi}}]{takasu2004}
\bibinfo{author}{\bibfnamefont{Y.}~\bibnamefont{Takasu}},
  \bibinfo{author}{\bibfnamefont{K.}~\bibnamefont{Komori}},
  \bibinfo{author}{\bibfnamefont{K.}~\bibnamefont{Honda}},
  \bibinfo{author}{\bibfnamefont{M.}~\bibnamefont{Kumakura}},
  \bibinfo{author}{\bibfnamefont{T.}~\bibnamefont{Yabuzaki}}, \bibnamefont{and}
  \bibinfo{author}{\bibfnamefont{Y.}~\bibnamefont{Takahashi}},
  \bibinfo{journal}{Phys. Rev. Lett.} \textbf{\bibinfo{volume}{93}},
  \bibinfo{pages}{123202} (\bibinfo{year}{2004}).

\bibitem[{\citenamefont{Tojo et~al.}(2006)\citenamefont{Tojo, Kitagawa,
  Enomoto, Kato, Takasu, Kumakura, and Takahashi}}]{tojo2006}
\bibinfo{author}{\bibfnamefont{S.}~\bibnamefont{Tojo}},
  \bibinfo{author}{\bibfnamefont{M.}~\bibnamefont{Kitagawa}},
  \bibinfo{author}{\bibfnamefont{K.}~\bibnamefont{Enomoto}},
  \bibinfo{author}{\bibfnamefont{Y.}~\bibnamefont{Kato}},
  \bibinfo{author}{\bibfnamefont{Y.}~\bibnamefont{Takasu}},
  \bibinfo{author}{\bibfnamefont{M.}~\bibnamefont{Kumakura}}, \bibnamefont{and}
  \bibinfo{author}{\bibfnamefont{Y.}~\bibnamefont{Takahashi}},
  \bibinfo{journal}{Phys. Rev. Lett.} \textbf{\bibinfo{volume}{96}},
  \bibinfo{pages}{153201} (\bibinfo{year}{2006}).

\bibitem[{\citenamefont{Mickelson et~al.}(2005)\citenamefont{Mickelson,
  Martinez, Saenz, Nagel, Chen, Killian, Pellegrini, and
  C\^ot\'{e}}}]{michelson2005}
\bibinfo{author}{\bibfnamefont{P.~G.} \bibnamefont{Mickelson}},
  \bibinfo{author}{\bibfnamefont{Y.~N.} \bibnamefont{Martinez}},
  \bibinfo{author}{\bibfnamefont{A.~D.} \bibnamefont{Saenz}},
  \bibinfo{author}{\bibfnamefont{S.~B.} \bibnamefont{Nagel}},
  \bibinfo{author}{\bibfnamefont{Y.~C.} \bibnamefont{Chen}},
  \bibinfo{author}{\bibfnamefont{T.~C.} \bibnamefont{Killian}},
  \bibinfo{author}{\bibfnamefont{P.}~\bibnamefont{Pellegrini}},
  \bibnamefont{and}
  \bibinfo{author}{\bibfnamefont{R.}~\bibnamefont{C\^ot\'{e}}},
  \bibinfo{journal}{Phys. Rev. Lett.} \textbf{\bibinfo{volume}{95}},
  \bibinfo{pages}{223002} (\bibinfo{year}{2005}).

\bibitem[{\citenamefont{Fedichev et~al.}(1996)\citenamefont{Fedichev, Kagan,
  Shlyapnikov, and M.Walraven}}]{fedichev1996}
\bibinfo{author}{\bibfnamefont{P.~O.} \bibnamefont{Fedichev}},
  \bibinfo{author}{\bibfnamefont{Y.}~\bibnamefont{Kagan}},
  \bibinfo{author}{\bibfnamefont{G.~V.} \bibnamefont{Shlyapnikov}},
  \bibnamefont{and} \bibinfo{author}{\bibfnamefont{J.~T.}
  \bibnamefont{M.Walraven}}, \bibinfo{journal}{Phys. Rev. Lett.}
  \textbf{\bibinfo{volume}{77}}, \bibinfo{pages}{2913} (\bibinfo{year}{1996}).

\bibitem[{\citenamefont{Ciurylo et~al.}(2005)\citenamefont{Ciurylo, Tiesinga,
  and Julienne}}]{ciurylo2005}
\bibinfo{author}{\bibfnamefont{R.}~\bibnamefont{Ciurylo}},
  \bibinfo{author}{\bibfnamefont{E.}~\bibnamefont{Tiesinga}}, \bibnamefont{and}
  \bibinfo{author}{\bibfnamefont{P.~S.} \bibnamefont{Julienne}},
  \bibinfo{journal}{Phys. Rev. A} \textbf{\bibinfo{volume}{71}},
  \bibinfo{pages}{030701} (\bibinfo{year}{2005}).

\bibitem[{\citenamefont{Ido and Katori}(2003)}]{ido2003}
\bibinfo{author}{\bibfnamefont{T.}~\bibnamefont{Ido}} \bibnamefont{and}
  \bibinfo{author}{\bibfnamefont{H.}~\bibnamefont{Katori}},
  \bibinfo{journal}{Phys. Rev. Lett.} \textbf{\bibinfo{volume}{91}},
  \bibinfo{pages}{053001} (\bibinfo{year}{2003}).

\bibitem[{\citenamefont{Takamoto and Katori}(2003)}]{takamoto2003}
\bibinfo{author}{\bibfnamefont{M.}~\bibnamefont{Takamoto}} \bibnamefont{and}
  \bibinfo{author}{\bibfnamefont{H.}~\bibnamefont{Katori}},
  \bibinfo{journal}{Phys. Rev. Lett.} \textbf{\bibinfo{volume}{91}},
  \bibinfo{pages}{223001} (\bibinfo{year}{2003}).

\bibitem[{\citenamefont{Zelevinsky et~al.}(2006)\citenamefont{Zelevinsky, Boyd,
  Ludlow, Ido, Ye, Ciurylo, Naidon, and Julienne}}]{zelevinsky2006}
\bibinfo{author}{\bibfnamefont{T.}~\bibnamefont{Zelevinsky}},
  \bibinfo{author}{\bibfnamefont{M.~M.} \bibnamefont{Boyd}},
  \bibinfo{author}{\bibfnamefont{A.~D.} \bibnamefont{Ludlow}},
  \bibinfo{author}{\bibfnamefont{T.}~\bibnamefont{Ido}},
  \bibinfo{author}{\bibfnamefont{J.}~\bibnamefont{Ye}},
  \bibinfo{author}{\bibfnamefont{R.}~\bibnamefont{Ciurylo}},
  \bibinfo{author}{\bibfnamefont{P.}~\bibnamefont{Naidon}}, \bibnamefont{and}
  \bibinfo{author}{\bibfnamefont{P.~S.} \bibnamefont{Julienne}},
  \bibinfo{journal}{Phys. Rev. Lett.} \textbf{\bibinfo{volume}{96}},
  \bibinfo{pages}{203201} (\bibinfo{year}{2006}).

\bibitem[{\citenamefont{Olshanii}(1998)}]{olshanii1998}
\bibinfo{author}{\bibfnamefont{M.}~\bibnamefont{Olshanii}},
  \bibinfo{journal}{Phys. Rev. Lett.} \textbf{\bibinfo{volume}{81}},
  \bibinfo{pages}{938} (\bibinfo{year}{1998}).

\bibitem[{\citenamefont{Petrov and Shlyapnikov}(2001)}]{petrov2001}
\bibinfo{author}{\bibfnamefont{D.~S.} \bibnamefont{Petrov}} \bibnamefont{and}
  \bibinfo{author}{\bibfnamefont{G.~V.} \bibnamefont{Shlyapnikov}},
  \bibinfo{journal}{Phys. Rev. A} \textbf{\bibinfo{volume}{64}},
  \bibinfo{pages}{012706} (\bibinfo{year}{2001}).

\bibitem[{\citenamefont{Naidon et~al.}(2006)\citenamefont{Naidon, Tiesinga,
  Mitchell, and Julienne}}]{naidon2006}
\bibinfo{author}{\bibfnamefont{P.}~\bibnamefont{Naidon}},
  \bibinfo{author}{\bibfnamefont{E.}~\bibnamefont{Tiesinga}},
  \bibinfo{author}{\bibfnamefont{W.~F.} \bibnamefont{Mitchell}},
  \bibnamefont{and} \bibinfo{author}{\bibfnamefont{P.~S.}
  \bibnamefont{Julienne}}, \bibinfo{journal}{physics/0607140}
  (\bibinfo{year}{2006}).

\bibitem[{\citenamefont{Wouters et~al.}(2003)\citenamefont{Wouters, Tempere,
  and Devreese}}]{wouter2003}
\bibinfo{author}{\bibfnamefont{M.}~\bibnamefont{Wouters}},
  \bibinfo{author}{\bibfnamefont{J.}~\bibnamefont{Tempere}}, \bibnamefont{and}
  \bibinfo{author}{\bibfnamefont{J.~T.} \bibnamefont{Devreese}},
  \bibinfo{journal}{Phys. Rev. A} \textbf{\bibinfo{volume}{68}},
  \bibinfo{pages}{053603} (\bibinfo{year}{2003}).

\bibitem[{\citenamefont{Granger and Blume}(2004)}]{granger2004}
\bibinfo{author}{\bibfnamefont{B.~E.} \bibnamefont{Granger}} \bibnamefont{and}
  \bibinfo{author}{\bibfnamefont{D.}~\bibnamefont{Blume}},
  \bibinfo{journal}{Phys. Rev. Lett.} \textbf{\bibinfo{volume}{92}},
  \bibinfo{pages}{133202} (\bibinfo{year}{2004}).

\bibitem[{\citenamefont{Yurovsky}(2005)}]{yurovsky2005}
\bibinfo{author}{\bibfnamefont{V.~A.} \bibnamefont{Yurovsky}},
  \bibinfo{journal}{Phys. Rev. A} \textbf{\bibinfo{volume}{71}},
  \bibinfo{pages}{012709} (\bibinfo{year}{2005}).

\bibitem[{\citenamefont{Yurovsky}(2006)}]{yurovsky2006}
\bibinfo{author}{\bibfnamefont{V.~A.} \bibnamefont{Yurovsky}},
  \bibinfo{journal}{Phys. Rev. A} \textbf{\bibinfo{volume}{73}},
  \bibinfo{pages}{052709} (\bibinfo{year}{2006}).

\bibitem[{\citenamefont{Bolda et~al.}(2003)\citenamefont{Bolda, Tiesinga, and
  Julienne}}]{bolda2003}
\bibinfo{author}{\bibfnamefont{E.~L.} \bibnamefont{Bolda}},
  \bibinfo{author}{\bibfnamefont{E.}~\bibnamefont{Tiesinga}}, \bibnamefont{and}
  \bibinfo{author}{\bibfnamefont{P.~S.} \bibnamefont{Julienne}},
  \bibinfo{journal}{Phys. Rev. A} \textbf{\bibinfo{volume}{68}},
  \bibinfo{pages}{032702} (\bibinfo{year}{2003}).

\bibitem[{\citenamefont{Kanjilal and Blume}(2004)}]{kanjilal2004}
\bibinfo{author}{\bibfnamefont{K.}~\bibnamefont{Kanjilal}} \bibnamefont{and}
  \bibinfo{author}{\bibfnamefont{D.}~\bibnamefont{Blume}},
  \bibinfo{journal}{Phys. Rev. A} \textbf{\bibinfo{volume}{70}},
  \bibinfo{pages}{042709} (\bibinfo{year}{2004}).

\bibitem[{\citenamefont{Stock et~al.}(2005)\citenamefont{Stock, Silberfarb,
  Bolda, and Deutsch}}]{stock2005}
\bibinfo{author}{\bibfnamefont{R.}~\bibnamefont{Stock}},
  \bibinfo{author}{\bibfnamefont{A.}~\bibnamefont{Silberfarb}},
  \bibinfo{author}{\bibfnamefont{E.~L.} \bibnamefont{Bolda}}, \bibnamefont{and}
  \bibinfo{author}{\bibfnamefont{I.~H.} \bibnamefont{Deutsch}},
  \bibinfo{journal}{Phys. Rev. Lett.} \textbf{\bibinfo{volume}{94}},
  \bibinfo{pages}{023202} (\bibinfo{year}{2005}).

\bibitem[{\citenamefont{Kanjilal and Blume}(2006)}]{kanjilal2006}
\bibinfo{author}{\bibfnamefont{K.}~\bibnamefont{Kanjilal}} \bibnamefont{and}
  \bibinfo{author}{\bibfnamefont{D.}~\bibnamefont{Blume}},
  \bibinfo{journal}{Phys. Rev. A} \textbf{\bibinfo{volume}{73}},
  \bibinfo{pages}{060701} (\bibinfo{year}{2006}).

\bibitem[{\citenamefont{Blume and Greene}(2002)}]{blume2002}
\bibinfo{author}{\bibfnamefont{D.}~\bibnamefont{Blume}} \bibnamefont{and}
  \bibinfo{author}{\bibfnamefont{C.~H.} \bibnamefont{Greene}},
  \bibinfo{journal}{Phys. Rev. A} \textbf{\bibinfo{volume}{65}},
  \bibinfo{pages}{043613} (\bibinfo{year}{2002}).

\bibitem[{\citenamefont{Bolda et~al.}(2002)\citenamefont{Bolda, Tiesinga, and
  Julienne}}]{bolda2002}
\bibinfo{author}{\bibfnamefont{E.~L.} \bibnamefont{Bolda}},
  \bibinfo{author}{\bibfnamefont{E.}~\bibnamefont{Tiesinga}}, \bibnamefont{and}
  \bibinfo{author}{\bibfnamefont{P.~S.} \bibnamefont{Julienne}},
  \bibinfo{journal}{Phys. Rev. A} \textbf{\bibinfo{volume}{66}},
  \bibinfo{pages}{013403} (\bibinfo{year}{2002}).

\bibitem[{\citenamefont{Bergeman et~al.}(2003)\citenamefont{Bergeman, Moore,
  and Olshanii}}]{bergeman2003a}
\bibinfo{author}{\bibfnamefont{T.}~\bibnamefont{Bergeman}},
  \bibinfo{author}{\bibfnamefont{M.~G.} \bibnamefont{Moore}}, \bibnamefont{and}
  \bibinfo{author}{\bibfnamefont{M.}~\bibnamefont{Olshanii}},
  \bibinfo{journal}{Phys. Rev. Lett.} \textbf{\bibinfo{volume}{91}},
  \bibinfo{pages}{163201} (\bibinfo{year}{2003}).

\bibitem[{\citenamefont{Moore et~al.}(2004)\citenamefont{Moore, Bergeman, and
  Olshanii}}]{moore2004}
\bibinfo{author}{\bibfnamefont{M.~G.} \bibnamefont{Moore}},
  \bibinfo{author}{\bibfnamefont{T.}~\bibnamefont{Bergeman}}, \bibnamefont{and}
  \bibinfo{author}{\bibfnamefont{M.}~\bibnamefont{Olshanii}},
  \bibinfo{journal}{J. Phys. IV France} \textbf{\bibinfo{volume}{116}},
  \bibinfo{pages}{69} (\bibinfo{year}{2004}), \bibinfo{note}{cond-mat/0210556}.

\bibitem[{\citenamefont{Bohn and Julienne}(1999)}]{bohn1999}
\bibinfo{author}{\bibfnamefont{J.~L.} \bibnamefont{Bohn}} \bibnamefont{and}
  \bibinfo{author}{\bibfnamefont{P.~S.} \bibnamefont{Julienne}},
  \bibinfo{journal}{Phys. Rev. A} \textbf{\bibinfo{volume}{60}},
  \bibinfo{pages}{414} (\bibinfo{year}{1999}).

\bibitem[{\citenamefont{Fioretti et~al.}(1998)\citenamefont{Fioretti, Comparat,
  Crubellier, Dulieu, F.Masnou-Seeuws, and Pillet}}]{fioretti1998}
\bibinfo{author}{\bibfnamefont{A.}~\bibnamefont{Fioretti}},
  \bibinfo{author}{\bibfnamefont{D.}~\bibnamefont{Comparat}},
  \bibinfo{author}{\bibfnamefont{A.}~\bibnamefont{Crubellier}},
  \bibinfo{author}{\bibfnamefont{O.}~\bibnamefont{Dulieu}},
  \bibinfo{author}{\bibnamefont{F.Masnou-Seeuws}}, \bibnamefont{and}
  \bibinfo{author}{\bibfnamefont{P.}~\bibnamefont{Pillet}},
  \bibinfo{journal}{Phys. Rev. Lett.} \textbf{\bibinfo{volume}{80}},
  \bibinfo{pages}{4402} (\bibinfo{year}{1998}).

\bibitem[{\citenamefont{Bethe}(1949)}]{bethe1949}
\bibinfo{author}{\bibfnamefont{H.~A.} \bibnamefont{Bethe}},
  \bibinfo{journal}{Phys. Rev.} \textbf{\bibinfo{volume}{76}},
  \bibinfo{pages}{38} (\bibinfo{year}{1949}).

\bibitem[{\citenamefont{Yasuda et~al.}(2006)\citenamefont{Yasuda, Kishimoto,
  Takamoto, and Katori}}]{yasuda2006}
\bibinfo{author}{\bibfnamefont{M.}~\bibnamefont{Yasuda}},
  \bibinfo{author}{\bibfnamefont{T.}~\bibnamefont{Kishimoto}},
  \bibinfo{author}{\bibfnamefont{M.}~\bibnamefont{Takamoto}}, \bibnamefont{and}
  \bibinfo{author}{\bibfnamefont{H.}~\bibnamefont{Katori}},
  \bibinfo{journal}{Phys. Rev. A} \textbf{\bibinfo{volume}{73}},
  \bibinfo{pages}{011403} (\bibinfo{year}{2006}).

\bibitem[{\citenamefont{Petrov et~al.}(2000)\citenamefont{Petrov, Holzmann, and
  Shlyapnikov}}]{petrov2000a}
\bibinfo{author}{\bibfnamefont{D.~S.} \bibnamefont{Petrov}},
  \bibinfo{author}{\bibfnamefont{M.}~\bibnamefont{Holzmann}}, \bibnamefont{and}
  \bibinfo{author}{\bibfnamefont{G.~V.} \bibnamefont{Shlyapnikov}},
  \bibinfo{journal}{Phys. Rev. Lett.} \textbf{\bibinfo{volume}{84}},
  \bibinfo{pages}{2551} (\bibinfo{year}{2000}).

\bibitem[{\citenamefont{Peano et~al.}(2005)\citenamefont{Peano, amd C.~Mora,
  and Egger}}]{peano2005}
\bibinfo{author}{\bibfnamefont{V.}~\bibnamefont{Peano}},
  \bibinfo{author}{\bibfnamefont{M.~T.} \bibnamefont{amd C.~Mora}},
  \bibnamefont{and} \bibinfo{author}{\bibfnamefont{R.}~\bibnamefont{Egger}},
  \bibinfo{journal}{New J. Phys.} \textbf{\bibinfo{volume}{7}},
  \bibinfo{pages}{192} (\bibinfo{year}{2005}).

\bibitem[{\citenamefont{Girardeau}(1960)}]{girardeau1960}
\bibinfo{author}{\bibfnamefont{M.}~\bibnamefont{Girardeau}},
  \bibinfo{journal}{J. Math. Phys.} \textbf{\bibinfo{volume}{1}},
  \bibinfo{pages}{516} (\bibinfo{year}{1960}).

\bibitem[{\citenamefont{Kinoshita et~al.}(2005)\citenamefont{Kinoshita, Wenger,
  and Weiss}}]{kinoshita2005}
\bibinfo{author}{\bibfnamefont{T.}~\bibnamefont{Kinoshita}},
  \bibinfo{author}{\bibfnamefont{T.}~\bibnamefont{Wenger}}, \bibnamefont{and}
  \bibinfo{author}{\bibfnamefont{D.~S.} \bibnamefont{Weiss}},
  \bibinfo{journal}{Phys. Rev. Lett.} \textbf{\bibinfo{volume}{95}},
  \bibinfo{pages}{190406} (\bibinfo{year}{2005}).

\bibitem[{\citenamefont{Olshanii and Dunjko}(2003)}]{olshanii2003}
\bibinfo{author}{\bibfnamefont{M.}~\bibnamefont{Olshanii}} \bibnamefont{and}
  \bibinfo{author}{\bibfnamefont{V.}~\bibnamefont{Dunjko}},
  \bibinfo{journal}{Phys. Rev. Lett.} \textbf{\bibinfo{volume}{91}},
  \bibinfo{pages}{090401} (\bibinfo{year}{2003}).

\bibitem[{\citenamefont{Gangardt and Shlyapnikov}(2003)}]{gangardt2003}
\bibinfo{author}{\bibfnamefont{D.}~\bibnamefont{Gangardt}} \bibnamefont{and}
  \bibinfo{author}{\bibfnamefont{G.}~\bibnamefont{Shlyapnikov}},
  \bibinfo{journal}{Phys. Rev. Lett.} \textbf{\bibinfo{volume}{90}},
  \bibinfo{pages}{010401} (\bibinfo{year}{2003}).

\bibitem[{\citenamefont{Ciurylo et~al.}(2006)\citenamefont{Ciurylo, Tiesinga,
  and Julienne}}]{ciurylo2006}
\bibinfo{author}{\bibfnamefont{R.}~\bibnamefont{Ciurylo}},
  \bibinfo{author}{\bibfnamefont{E.}~\bibnamefont{Tiesinga}}, \bibnamefont{and}
  \bibinfo{author}{\bibfnamefont{P.~S.} \bibnamefont{Julienne}},
  \bibinfo{journal}{Phys. Rev. A} \textbf{\bibinfo{volume}{74}},
  \bibinfo{pages}{022710} (\bibinfo{year}{2006}).

\end{thebibliography}

\end{document}